\documentclass[review]{elsarticle}
 \pdfoutput=1
\usepackage{lineno,hyperref}
\usepackage{graphicx}
\usepackage{epstopdf}
\usepackage{setspace}
\usepackage{fancyhdr}
\usepackage{multirow}
\modulolinenumbers[5]

\journal{xxxx}

\begin{document}

\begin{frontmatter}

\title{Predicting intraday jumps in stock prices using liquidity measures and technical indicators}

\author{Ao Kong*}
\address{School of Finance, Nanjing University of Finance and Economics, Nanjing 210023, China}
\author{Hongliang Zhu}
\address{School of Management and Engineering, Nanjing University, Nanjing 210023, China}
\author{Robert Azencott}
\address{Department of Mathematics, University of Houston, TX 77204, USA}

\begin{abstract}
Predicting the intraday stock jumps is a significant but challenging problem in finance. Due to the instantaneity and imperceptibility characteristics of intraday stock jumps, relevant studies on their predictability remain limited. This paper proposes a data-driven approach to predict intraday stock jumps using the information embedded in liquidity measures and technical indicators. Specifically, a trading day is divided into a series of 5-minute intervals, and at the end of each interval, the candidate attributes defined by liquidity measures and technical indicators are input into machine learning algorithms to predict the arrival of a stock jump as well as its direction in the following 5-minute interval. Empirical study is conducted on the level-2 high-frequency data of 1271 stocks in the Shenzhen Stock Exchange of China to validate our approach. The result provides initial evidence of the predictability of jump arrivals and jump directions using level-2 stock data as well as the effectiveness of using a combination of liquidity measures and technical indicators in this prediction. We also reveal the superiority of using random forest compared to other machine learning algorithms in building prediction models. Importantly, our study provides a portable data-driven approach that exploits liquidity and technical information from level-2 stock data to predict intraday price jumps of individual stocks.
\end{abstract}


\end{frontmatter}

\section{Introduction}

Stock price jumps refer to large discontinuous changes in the path of stock returns and are mainly associated with company-specific events, market news announcements or large arbitrage hedging activities \citep{Merton1976, Lee2008,Boudt2014}. Compared with continuous price changes, stock jumps bring higher risk, leading to different applications in risk management, option pricing and asset allocation. While various studies of stock jumps, such as jump detection, causes of jumps, jump modelling and the effect of jumps on the market, have attracted much attention\citep{Lee2008, Evans2011, Lee2012, Boudt2014, Huang2018, Ma2019}, research on the predictability of stock jumps, especially under the intraday scenario, is very limited. This limitation might be because stock jumps normally occur instantly with imperceptible precursors, which makes it very challenging to capture the predictive features before their arrival. This study thus attempts to complement the literature in this area by designing an approach to predict intraday stock jumps and testing the method on thousands of stocks in a Chinese stock market.

To design a mechanism for jump prediction, statistically significant jumps in stock prices must be detected accurately. Most jump identification methods are designed for low-frequency data, which can only identify the existence of jumps within a day. Recent development of volatility modelling with high-frequency price series promotes advanced methods to detect the intraday timing of jump arrivals\citep{Lee2008, Boudt2011}, which enables the analysis of the instant ex ante transaction behaviour of stocks.

Based on these intraday jump detection algorithms, some studies have investigated the intraday liquidity dynamics and found significant abnormal movements before stock jumps\citep{Boudt2014, Wan2017, Bedowska2016}. Although some of the liquidity measures have proved to be associated with the occurrence of stock jumps, they perform poorly in predicting stock jumps individually. While all these studies concentrated on only single variables, the question of whether combinations of liquidity measures have higher explanatory power in predicting stock jumps remains open.

Different from liquidity measures, technical indicators, originating from technical analysis for stock investing, characterize the market situation in an alternative way: a liquidity measure describes the market quality at the current time, whereas a technical indicator studies the market movement within a comparably longer time window. Although initially designed for analysing low-frequency stock data, technical indicators have been successfully used to predict high-frequency stock price series in recent literature\citep{Son2012, Sun2019, Borovkova2019}. However, to the best of our knowledge, no study has been performed to analyse their predictive power for stock jumps.

Therefore, this study intends to combine both liquidity measures and technical indicators to predict intraday stock jumps. Our proposed approach is based on level-2 high-frequency data, which are ultra informative for tracking the instant movement of these attributes before stock jumps. We divide a whole trading day into a series of 5-minute intervals, and at the end of each interval, our model determines whether a stock jump will occur in the next 5 minutes. Ten types of liquidity measures and eighteen types of technical indicators are adopted to define hundreds of candidate attributes that characterize the current and prior market dynamics of each interval. Since the predictive power of a single attribute is very weak, various combinations need to be analysed with a large quantity of data to search for potential important patterns indicating the arrival of stock jumps. Our prediction model is built and tested universally on the data of 1271 stocks in Chinese stock exchange to avoid overfitting issues. Such a task is too complicated for a human observer; hence, machine learning is used in our prediction owing to the advances in addressing high-dimensional datasets and complex modelling problems.

The main contributions of our work are as follows:
\begin{enumerate}[1)]
\item We propose an approach to predict intraday stock jumps using level-2 high-frequency data. To the best of our knowledge, only two studies \citep{Makinen2019, Zheng2013} consider the prediction of individual stock jumps. While \citet{Zheng2013} focused on two specific types of jumps, called inter-trade and trade-through price jumps, we concentrate on a more general price jump defined by \citet{Merton1976}. \cite{Makinen2019} studied the same price jump as us, but used tick-by-tick data and focused only on five individual stock. In some countries such as China, tick data is not available to the public. Besides, tick data is substantially affected by market microstructure noise and should be used cautiously\citep{Jiang2011}. As a result, our methodology is based on the level-2 high-frequency data of 1271 stocks in China and relies on a different set of processing algorithms.

\item Our approach combines the explanatory power of liquidity measures and technical indicators to enhance intraday stock jump prediction. On one hand, liquidity measures have been examined only individually concerning their predictive power for price jumps, while combinations of them have been ignored; on the other hand, no technical indicator has been studied regarding the occurrence of jumps, given that such indicators have been successfully used to predict stock prices in existing literatures.

\item Apart from predicting the arrival of intraday stock jumps, our study sheds additional light on the predictability of jump directions. \cite{Makinen2019} have studied a similar problem, but obtained pessimistic results, while we proved in this study that the jump direction can be predicted with better result using our constructed features and model construction scheme.
\end{enumerate}

\section{Literature Review}
\subsection{Price jump detection}

Early studies on price jumps normally use daily data and treat volatility and its jump component as unobservable hidden variables that are modelled by GARCH-jump or SV-jump models. Such methods involve uncertain pre-setting of parameter forms and complex parameter estimation, which obstructs the accurate estimation of price jumps. With the accumulation of high-frequency data, \citet{Andersen1998}, \citet{Andersen2001} and \citet{Barndorff2002} introduced "estimated volatility" calculated from the sum of squared high-frequency returns, thereby explicitly representing and modelling financial volatility. \citeauthor{Barndorff2004}(\citeyear{Barndorff2004}, \citeyear{Barndorff2006}) further used the difference between the realized volatility and the bipower variation as an unbiased estimator of discrete jump variance, giving a nonparametric test (BNS test) for jump detection. Such a nonparametric method avoids complex parameter setting, leading to higher jump detection efficiency. Since then, the BNS method has been further improved using various forms of variation and volatility \citep{Huang2005,Jiang2008,Jacod2009,Ait-Sahalia2009,Mancini2009,Corsi2010,Podolskij2010,Andersen2012}.

However, the above BNS type methods can only determine whether a jump occurs within a trading day, without specifying the time and number of the jumps. To solve this problem, \citet{Andersen2007a} designed a test (ABD test) for intraday jump detection that compares the standardized intraday returns to critical values from the normal distribution. However, \citet{Lee2008} found that the ABD test is too permissive and proposed a more general test (LM test) using critical values from the limit distribution of the maximal test statistic. \citet{Boudt2011} further demonstrated that the accuracy and robustness of the LM test can be improved by considering the intra-week periodicity in volatility. These methods are normally applied to 5-minute interval data. Alternatively, \citet{Lee2012} and \citet{Christensen2014} proposed two methods to identify price jumps from ultra high-frequency data at tick frequency since they believe identification is aided by the finest resolution of price data. The spurious detection by their methods was further eliminated by \citet{Bajgrowicz2016} using an explicit thresholding technique.

The finest tick-by-tick resolution preserves sufficient information but induces higher market microstructure noise, which biases the jump detection. Although determining the optimal sampling frequency for jump detection is an ongoing research topic, \citet{Liu2015} found little evidence that the consensus 5-minute interval is surpassed by any other choice of sampling frequency in terms of realized volatility estimation. A similar argument was made by \citet{Jiang2011} that tick data should be used cautiously because of concerns about market microstructure effects. Therefore, we base our jump detection procedure on the consensus 5-minute sampled data.

\subsection{Liquidity dynamics before price jumps}
Individual stock jumps are associated with not only macroeconomic events but also sudden firm-specific news \citep{Lahaye2011,Bollerslev2008,Lee2008}. In addition, recent research by \citet{Boudt2014} and \citet{Bedowska2016} found that pure liquidity variation might be the main reason for a large majority of individual stock jumps. Regardless of the causes of jumps, scientists tend to discover clues before the arrival of stock jumps from a market microstructure perspective since macro variables appear to have weak forecasting power \citep{Jiang2011, Bedowska2016}.

\citet{Boudt2014} found a substantial phenomenon of abnormal liquidity dynamics before stock jumps and noted that some liquidity measures, such as order imbalance and depth imbalance, are informative for price discovery, especially after the arrival of news. Similar results in the T-bond market were obtained by \citet{Jiang2011}, who compared the capacity of macroeconomic news announcements versus liquidity shocks to explain price jumps and found that the latter has more explanatory power, even during the preannouncement period. \citet{Bedowska2016} analysed the pattern of liquidity measures around price jumps on the Warsaw Stock Exchange and summarized that jumps are accompanied by abnormalities in some liquidity variables, such as trading volume, number of transactions, volatility, and market depth.

While the majority of works is devoted to the analysis of developed markets, \citet{Wan2017} focused on the Chinese stock market and observed, based on a large number of jumps, that liquidity measures show abnormal patterns around jumps. Moreover, in contrast to other developed countries, the Chinese government does not release new data and policies at a fixed time interval, and no data suppliers provide estimations of news expectations and surprises to the public. Thus, in the Chinese stock market, prices may gradually adjust to news announcements. As in other similar markets with random news announcement dates, we may therefore conjecture that clues about jumps in liquidity measures may be more significant than those in developed markets. Although some liquidity measures are found to be abnormal before price jumps, it remains unclear whether combinations of liquidity variables have higher prediction power, despite the weak power of individual variables.

\subsection{Technical indicators in high-frequency data prediction}
Technical indicators, developed for over a century, are commonly used analysis tools to determine the market trend of an asset using historical price and volume information. Although technical indicators were originally designed for low-frequency data, typically daily data, a number of efforts have been made to explore their capacity in forecasting high-frequency intraday financial data. Notably, technical indicators contribute to intraday stock trend prediction whenever the sampling frequency is by tick\citep{Tanaka2007}, minutes\citep{Son2012} or hours\citep{Gunduz2017}.

\citet{Tanaka2007} examined the effectiveness of several frequently used technical indicators, such as momentum, moving average and relative strength index, for several-tick-ahead trend prediction of NYSE stock prices. \citet{Son2012} used eight types of technical indicators, including moving average, price rate of change, and relative strength index, to predict the trend direction of KOSPI200 index 5 minutes ahead. \citet{Gunduz2017} adopted more than 20 types of technical indicators to predict the hourly movements of 100 stocks in Borsa Istanbul. In general, \citet{Schulmeister2009} demonstrated, over a long period of time, that the profitability of technical indicators has shifted from daily to intraday as a result of adaption to the increasing speed of stock transactions.

The related research is not limited to stock markets. In the work of \citet{Nakano2018}, \citet{Ozturk2016} and \citet{Lee2003}, the profitability of using technical indicators for intraday price prediction is demonstrated in bitcoin, foreign exchange and index futures markets. Apart from trend prediction, technical indicators based on high-frequency data are also successfully applied in various other fields, such as determining optimal trading time\citep{Cervello2015}, ranking stocks\citep{Wang2018} and predicting market shocks\citep{Sun2019}.

Intraday jump is an important component of high-frequency asset series. Despite achieving satisfactory results in high-frequency data prediction, technical indicators are rarely used in the analysis of intraday jumps extracted from high-frequency time series.

\section{Preliminaries} \label{background}

\subsection{Jump detection technique}\label{detection}
Our intraday jump detection phase is based on the LM test developed by \citet{Lee2008}. To improve the performance, we combine this technique with the robust volatility estimation methods proposed by \citet{Andersen2012} and \citet{Boudt2011}, as in the work of \citep{Wan2017}. The choice of working with 5-minute return series is consistent with many influential studies \citep[e.g.]{Lee2008, Jiang2011, Boudt2014}. Details of the technique are provided in the following.

A trading day is divided into $n$ of 5-minute intervals, and intraday jumps within each interval are detected. Let $r_{t,i}$ be the log return in the $i$th interval of day $t$, and let $\sigma^2_t$ be the integrated volatility on day $t$. \citet{Lee2008} estimated $\sigma^2_t$ by the realized bipower variation of the intraday returns
\begin{equation}
\hat{\sigma}^2_t =\frac{1}{n-1}\sum_{i=2}^n|r_{t,i-1}||r_{t,i}|,
\end{equation}
and the jump detecting statistic is constructed by
\begin{equation}
L_{t,i} = \frac{r_{t,i}}{\hat{\sigma}_t}.
\end{equation}

To improve the robustness of volatility and jump detection, \citet{Andersen2012} proposed the MedRV estimator $\hat{\sigma}^{med}_t$ to estimate the integrated volatility
\begin{equation}\hat{\sigma}^{med}_t = \sqrt{\frac{1}{nK-2}\frac{\pi}{6-4\sqrt{3}+\pi}\left(\frac{n}{nk-2}\right)\times \sum_{i=3}^{nk}median\{|\tilde{r}_{i-2}|, |\tilde{r}_{i-1}|, |\tilde{r}_i|\}},
\end{equation}
where $\{\tilde{r}_i, i=1,2,\cdots,nK\}$ are the return series of all the intervals in recent $K$ days. Here, we choose $K=5$ as in \cite{Wan2017}. In addition, the weighted standard deviation (WSD) estimator $f^{WSD}_{t,i}$ of $f_{t,i}$, suggested by \cite{Boudt2011}, is included as a periodic component to account for the intraweek periodicity in high-frequency returns. Then, the jump detecting statistic $L_{t,i}$ is replaced by
\begin{equation}
\tilde{L}_{t,i} = \frac{r_{t,i}}{\hat{\sigma}^{med}_tf^{WSD}_{t,i}}.
\end{equation}

The final detected jump in the $i$th interval of day $t$ should satisfy
\begin{equation}
\frac{\tilde{L}_{t,i}-C_{nT}}{S_{nT}}>-\log(-\log(1-\alpha)),
\end{equation}
where $C_{nT}$ and $S_{nT}$ are defined by \citet{Lee2008} and $T$ is the total number of days.

\subsection{Liquidity measures}\label{lm}
Ten liquidity measures are utilized to describe the return, volatility, market width and market depth of a stock's trading in each 5-minute interval. Let $LM$ be any liquidity measure; then, its value computed in the $i$th interval of day $t$ is represented by $\{LM_{t,i}, i =1, 2, \cdots, n, t = 1,2,\cdots, T\}$, where $T$ is the total number of trading days. Table \ref{tab:lm} summarizes the ten liquidity measures.

\begin{table}[htbp]
\centering
\caption{Liquidity measures}\label{tab:lm}
\begin{tabular}{lll}
\hline\noalign{\smallskip}
Index     &Name        &Notation\\
\noalign{\smallskip}\hline\noalign{\smallskip}
1   &Return                      &$r_{t,i}$  \\
2   &Cumulative return           &$R_{t,i}$\\
3   &Number of trades            &$k_{t,i}$\\
4   &Trading volume              &$v_{t,i}$\\
5   &Trading size                &$s_{t,i}$\\
6   &Order imbalance             &$oi_{t,i}$\\
7   &Depth imbalance             &$di_{t,i}$\\
8  &Quoted spread               &$qs_{t,i}$\\
9  &Effective spread            &$es_{t,i}$\\
10  &Realized volatility         &$rv_{t,i}$\\
\noalign{\smallskip}\hline
\end{tabular}
\end{table}

Detailed computation of these measures is illustrated in Appendix A. These measures need standardization to be comparable between different stocks. Similar to the study of \citet{Boudt2014}, $k_{t,i}$, $v_{t,i}$ and $s_{t,i}$ are standardized by dividing by the median of their data at the same time of the previous 60 days and subtracting 1. The rest are standardized by subtracting the median of their data at the same time of the previous 60 days.

\subsection{Technical indicators}\label{ti}
Eighteen types of technical indicators are used to characterize the movement of a stock's price within a prior window. Technical indicators are usually functions of the open price, highest price, lowest price, close price and volume. In contrast to daily computation, in the context of our study, this price and volume information is extracted within each 5-minute interval. The computation of some indicators relies on a few lagged periods, hence the lagged intervals. Suppose there are $T$ trading days in total; then, there should be $nT$ intervals. Let $q$ denote the number of lagged intervals. Then, the list of any intraday technical indicator $TI$ over all the intervals can be represented by $\{TI(q)_k, k=1,2,\cdots,nT\}$. All the indicators are summarized in Table \ref{tab:iti}.

\begin{table}[htbp]
\centering
\caption{Intraday technical indicators}\label{tab:iti}
\begin{tabular}{lll}
\hline\noalign{\smallskip}
Index       &Name        &Notation            \\
\noalign{\smallskip}\hline\noalign{\smallskip}
1       &Price rate of change        &$PROC(q)_k$\\
2       &Volume rate of change          &$VROC(q)_k$\\
3       &Moving average of price       &$MA(q)_k$  \\
4       &Exponential moving average of price &$EMA(q)_k$\\
5       &Bias to MA           &$BIAS(q)_k$\\
6       &Bias to EMA            &$EBIAS(q)_k$\\
7       &Price oscillator to MA &$OSCP(q)_k$\\
8       &Price oscillator to EMA &$EOSCP(q)_k$\\
9      &Fast stochastic \%K     &$fK(q)_k$\\
10      &Fast stochastic \%D     &$fD(q)_k$\\
11      &Slow stochastic \%D      &$sD(q)_k$\\
12      &Commodity channel index         &$CCI(q)_k$\\
13      &Accumulation/Distribution oscillator  &$ADO_k$\\
14      &True range            &$TR_k$\\
15      &Price and volume trend              &$PVT_k$\\
16      &On balance volume         &$OBV_k$\\
17      &Negative volume index     &$NVI_k$\\
18      &Positive volume index       &$PVI_k$\\
\noalign{\smallskip}\hline
\end{tabular}
\end{table}

Detailed computation of these indicators is provided in Appendix B. These indicators also need to be standardized. $MA(q)_k$, $EMA(q)_k$ $TR_k$, $OBV_k$, $PVT_k$, $NVI_k$ and $PVI_k$ are standardized by dividing by the median of their data at the same time of the previous 60 days and subtracting 1. The rest are already ratios, so they are standardized by subtracting the median of their data at the same time of the previous 60 days.

\subsection{Class balancing techniques}
Most standard algorithms assume balanced class distributions. But for the jump prediction problem, the number of non-jumping instances is substantially larger than the number of jumping instances. Besides, the upward-jumping and downward-jumping class are severely imbalanced. Thus to ensure efficient feature selection and model construction, class balancing of the training data is required\citep{Veganzones2018,He2009}.

Two class balancing techniques, random undersampling\citep{He2009} and synthetic minority oversampling technique (SMOTE)\citep{Chawla2002}, are used in our study. Random undersampling technique simply randomly delete instances from the majority class. Synthetic minority oversampling technique (SMOTE), on the other hand, creates artificial data to add to the minority class. Namely, for each instance $x$ in the minority and some specified integer $K$, SMOTE first selects one of the K-nearest neighbors of $x$, say $\hat{x}$, whose Euclidian distances between $x$ and itself is the smallest. Then it adds $\delta(\hat{x}-x)$ to $x$, where $\delta\in [0,1]$ is a random number, to create an artificial instance.

The random undersampling is used to balance the jumping and non-jumping classes in a binary classification problem and both of the two techniques are utilized to balance the non-jumping,upward-jumping and downward-jumping classes in a trinary classification problem.

\subsection{Feature selection based on mutual information}\label{mRMR}
Mutual information, defined by \cite{Shannon2001}, is used to evaluate the mutual dependence of two random variables under the framework of information theory. The mutual information of random variables $X$ and $Y$ is estimated by
\begin{equation}\label{MI}
I(X,Y) = \frac{1}{N}\sum_{i=1}^N\ln\frac{p_{XY}(x_i, y_i)}{p_X(x_i)p_Y(y_i)},
\end{equation}
where $p_X(x_i), p_Y(y_i)$ and $p_{XY}(x_i,y_i)$ are the sample estimates of the marginal and joint density functions of random variables $X$ and $Y$. High mutual information indicates large dependency.

Mutual information provides a rank of all the candidate attributes in terms of their correlation with the target variable but cannot be used to select the optimal set of features. An optimal set of features should be highly informative but minimally redundant. In our study, the minimum redundancy-maximum relevance (mRMR) feature selection algorithm is applied in combination with a forward selection search strategy to select efficient sets of features based on mutual information \cite{Meyer2008}.

Given a target variable $y$ and all the candidate attributes $\{x_i\}_{i=1}^{d}$, in each step of mRMR, let $F_S$ be the set of already selected features. The relevance of an attribute $x_i\in F_{-S}$, where $F_{-S}$ is the set of not selected attributes, with the target variable $y$ is evaluated in terms of their mutual information $I(x_i,y)$, and the redundancy of $x_i$ with $F_S$ is evaluated by
\begin{equation}
RD_i = \sum_{x_j\in F_S}I(x_i, x_j)/|F_S|.
\end{equation}
Then, mRMR ranks all the attributes $x_i\in F_{-S}$ by the criterion
\begin{equation}
Cr(x_i) = I(x_i,y)-RD_i,
\end{equation}
and augments $F_S$ with the attribute that maximizes $Cr$. Therefore, for a given number $l$, the mRMR algorithm provides an optimal set of $l$ features after $l-1$ iteration steps.

\subsection{Model evaluation metrics}\label{eval}
The accuracy and F-measure are used to evaluate the performance of our jump prediction model. Accuracy is the most commonly used evaluation metric for a learning model. However, in our study, the correct prediction of a jump is more important than the correct prediction of a non-jump, so we incorporate the F-measure to evaluate specifically the prediction accuracy of jump instances.

In terms of a binary and a trinary classification problems, accuracy and F-measure are defined based on the counts in two types of confusion matrices as shown in Table \ref{tab:cm1} and Table \ref{tab:cm2}.

\begin{table}[htbp]
\centering
\caption{Confusion matrix for binary classification}\label{tab:cm1}
\begin{tabular}{cccccc}
\hline\noalign{\smallskip}
Actual$\downarrow$/Predicted$\rightarrow$        &Jump        &No jump          \\
\noalign{\smallskip}\hline\noalign{\smallskip}
Jump          &a        &b           \\
No jump        &c         &d          \\
\noalign{\smallskip}\hline
\end{tabular}
\end{table}

\begin{table}[htbp]
\centering
\caption{Confusion matrix for trinary classification}\label{tab:cm2}
\begin{tabular}{cccccccc}
\hline\noalign{\smallskip}
Actual$\downarrow$/Predicted$\rightarrow$        &Upward jump      &Downward jump  &No jump            \\
\noalign{\smallskip}\hline\noalign{\smallskip}
Upward jump          &$\tilde{a}$        &$\tilde{b}$     &$\tilde{c}$         \\
Downward jump       &$\tilde{d}$         &$\tilde{e}$      &$\tilde{f}$\\
No jump        &$\tilde{g}$        &$\tilde{h}$         &$\tilde{k}$\\
\noalign{\smallskip}\hline
\end{tabular}
\end{table}

Accuracy is simply the percentage of correctly predicted instances, and is computed by
\begin{equation}
acc = \frac{a+d}{a+b+c+d},
\end{equation}
and
\begin{equation}
acc = \frac{\tilde{a}+\tilde{e}+\tilde{k}}{\tilde{a}+\tilde{b}+\tilde{c}+\tilde{d}+\tilde{e}+\tilde{f}+\tilde{g}+\tilde{h}+\tilde{k}}
\end{equation}
respectively for Table \ref{tab:cm1} and Table \ref{tab:cm2}.

The F-measure, defined as the harmonic mean of precision and recall, is calculated by
\begin{equation}
fm = \frac{2\times precision\times recall}{precision+recall},
\end{equation}
where precision and recall are the ratio of correctly predicted jumping instances to the total predicted jumping instances and to the total actual jumping instances. For Table \ref{tab:cm1}, precision and recall are simply defined as
\[precision = \frac{a}{a+c}\quad and \quad recall = \frac{a}{a+b}.\]
For Table \ref{tab:cm2}, precision and recall are defined for the upward and downward jumping instances respectively to evaluate the prediction performance on the two types of jumps:
\[precision^+ = \frac{\tilde{a}}{\tilde{a}+\tilde{d}+\tilde{g}}\quad and \quad recall^+ = \frac{\tilde{a}}{\tilde{a}+\tilde{b}+\tilde{c}}\] for the upward jumping instances
, and
\[precision^- = \frac{\tilde{e}}{\tilde{b}+\tilde{e}+\tilde{h}}\quad and \quad recall^- = \frac{\tilde{e}}{\tilde{d}+\tilde{e}+\tilde{f}}\] for the downward jumping instances.

\section{Experimental Design}\label{design}
With the purpose of jump prediction, our study attempts to answer two questions: a) if a stock jump is arriving in the following five-minute interval; b) if an upward jump or a downward jump is arriving in the following five-minute interval. To answer the first question, a binary classification is needed to differentiate between jumping and non-jumping instances regardless of the jump directions. For the second question, a trinary classification problem should be implemented to distinguish among upward jumping, downward jumping and non-jumping instances. Whenever the prediction task is, we follow the similar framework shown in Figure \ref{framework}. Three phases are included in the framework, details of which are presented in the following.
\begin{figure}[htbp]
\begin{center}
\includegraphics[scale=0.7]{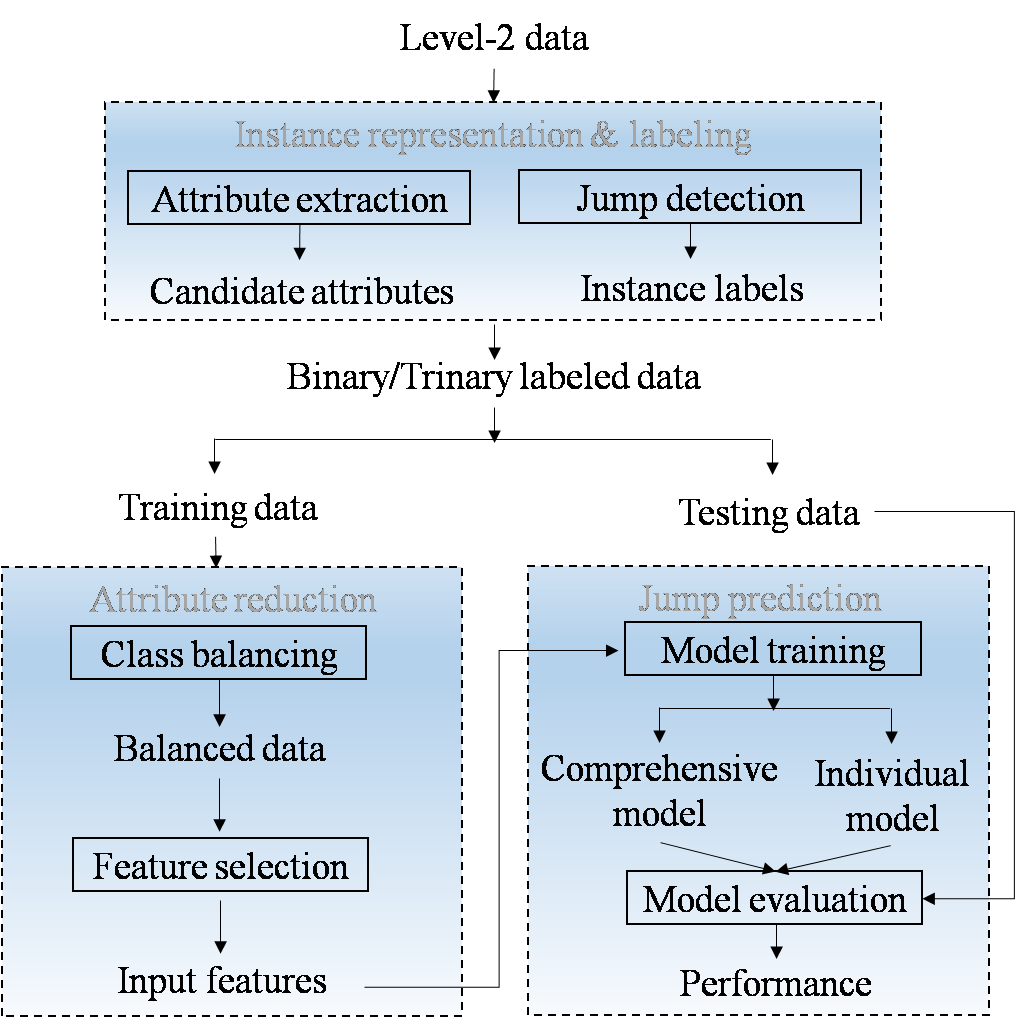}
\caption{Proposed design framework}\label{framework.png}
\end{center}
\end{figure}

\subsection{Instance representation and labelling}
\begin{figure}[htbp]
\begin{center}
\includegraphics[scale=0.46]{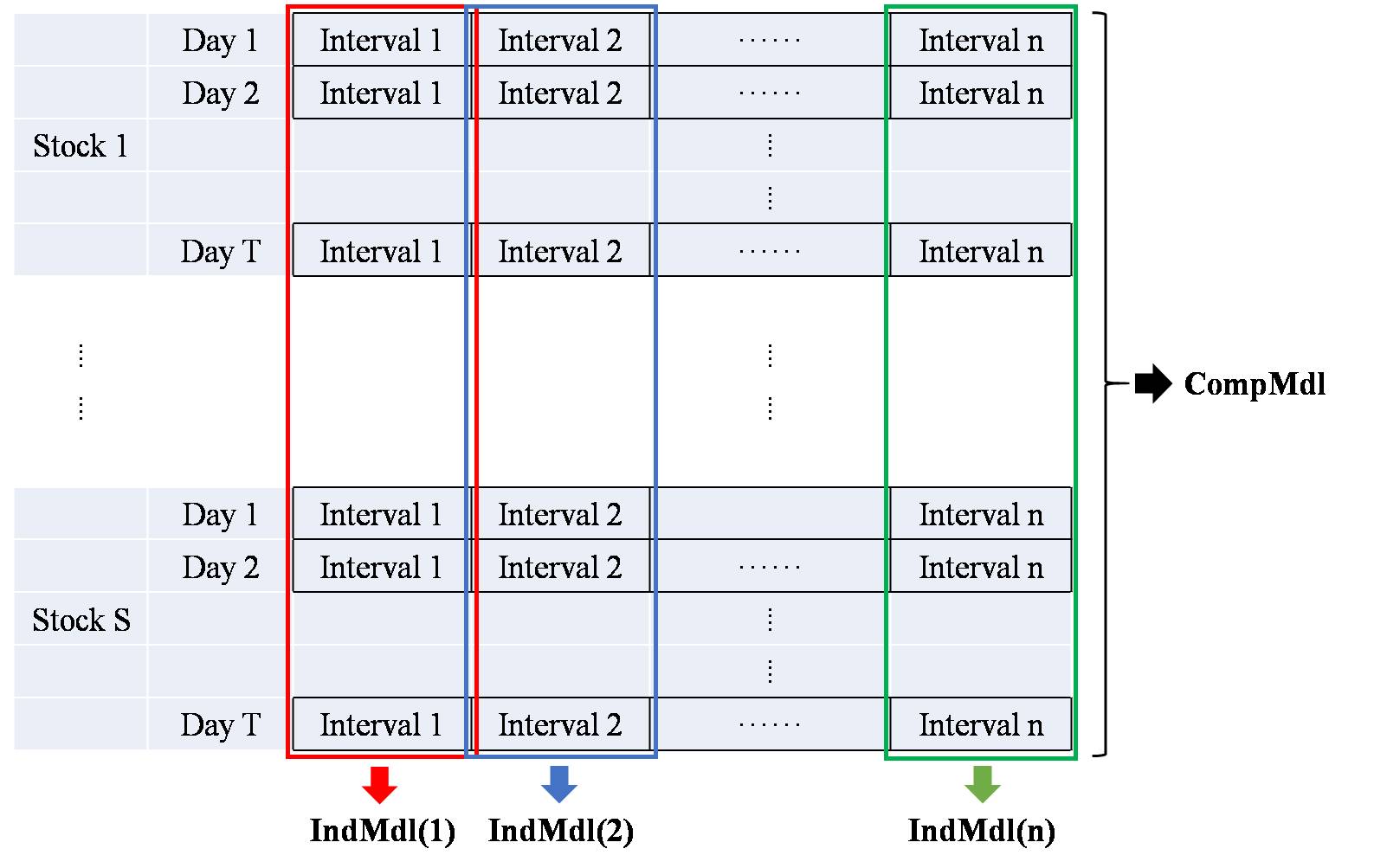}
\caption{Terminology of model building in our approach}\label{Fig:comp_ind_model}
\end{center}
\end{figure}

To forecast the occurence of price jumps every five minutes using information embedded in liquidity measures and technical indicators, we divided the market hours in a trading day into $n$ five-minute intervals. The liquidity measures and technical indicators are then computed every five minutes, yielding $n$ observations per stock. Summing over $T$ trading days of $S$ stocks, the whole dataset includes $n\times T\times S$ instances. Our approach intends to obtain a prediction model that can be used universally to all the stocks in the market, thus we use all the $n\times T\times S$ instances to train and test a comprehensive model $CompMdl$. In addition, considering that stock jumps exhibit time-varying characteristic which might follow a time-varying prediction model, we also construct individual prediction models from interval to interval attempting to explore the time-varying predictability of stock jumps. That is, we use the $T\times S$ instances in each interval to build an individual model $IndMdl(i), i = 1, 2, \cdots, n$. This terminology is explained more explicitly in Figure \ref{Fig:comp_ind_model}.

The jump events of each stock are detected by the technique described in section \ref{detection}. For the binary classification problem, where the jump direction is ignored, an instance is labeled by 1 or 0, where 1 denotes the arrival of a jump and 0 denotes continuous price change without jumps. For the trinary classification, the jumping instances are further divided into upward jumping and downward jumping instances and are labeled respectively by 1 and -1.

Because of the ¡¯curse of dimensionality¡¯, it is not favourable to select a large number of input attributes. Referring to the existing studies, such as \cite{Boudt2014} and \cite{Wan2017}, liquidity measures within a prior one-hour window, i.e., 12 intervals, are adopted as candidate attributes. We also consider the average value of each measure over the 12 intervals as a new attribute. Thus, we construct 130 candidate attributes from the 10 liquidity measures. No gold standard exists to choose the time lag for intraday technical indicators. To reflect the market change within a moderately longer period, lags of $q=5, 10, 20, 30$ are used to define our intraday technical indicators, which provides 54 candidate attributes. Thus, the total number of candidate attributes is 184.

Following the general scheme of machine learning, these processed data, of which each instance is represented by an attribute vector and labelled with 1/0 or 1/-1/0, are divided into training and testing sets. The training set is used in the model building phase to train a prediction model $CompMdl$, and the testing set is adopted for model validation. To build individual models $IndMdl(i)(i = 1,2,\cdots,n)$, the training and testing sets are further divided respectively into $n$ sets corresponding to $n$ intervals.

\subsection{Attribute reduction}

\subsubsection{Class balancing}\label{balance}
Assume that the sizes of the upward jumping, downward jumping and non-jumping classes in a training set are $M, N, L$. For the binary classification problem between jumping and non-jumping instances, the upward and downward jumping instances are grouped as one class with size $M+N$, and the other class consists $L$ non-jumping instances. Since $L$ is usually hundreds of times larger $M+N$, we simply adopt a random undersampling procedure on the non-jumping class to make its size equal $M+N$.

The second type of prediction is a trinary classification problem differentiating among the upward-jumping, downward-jumping and non-jumping instances. It is found that among all the jumping instances, the upward and downward ones are also imbalanced. Thus for this classification problem, we first use SMOTE technique to oversampling the minority between the downward and upward jumping sets to balance the two classes. Then we apply a random undersampling procedure as above to reduce the size of non-jumping class. After the two sampling steps, the sizes of the three classes should equal $\max(M,N)$. Here, SMOTE is used instead of random oversampling because it is a more powerful oversampling method to improve model performance and to avoid overfitting issues.

After balancing the number of instances among classes, a training set with balanced classes is processed to build a prediction model $Mdl$, where $Mdl$ can be a $CompMdl$ or a $IndMdl(i)$, $i = 1, 2, \cdots, n$. Taking into account the variation caused by undersampling or oversampling, 50 times of such class balancing procedure are conducted to construct 50 replicate training sets $repTr_j, j = 1, 2, \cdots 50$. Then the following feature selection and model training procedures are repeated on the 50 training sets to produce 50 replicate models $repMdl_j, j = 1, 2, \cdots, 50$. The size of each of the $repTr_j$ should be $2(M+N)$ for the binary classification problem and $3\max(N,N)$ for the trinary classification problem.

\subsubsection{Feature selection} \label{ftselect}
Efficient prediction requires the selection of a smaller subset of informative features from the 184 candidate attributes. Mutual information is used to evaluate the informative level of a feature to predict the target variable. Since we have to compare the mutual information between the candidate attributes, we shall be ascertain that the result is not blurred by their individual distributions. Thus, an adaptive partition technique instead of the commonly used equal-length partition to the axes is adopted during the estimation of mutual information (see \cite{Steuer2002} for details). Then, a feature selection procedure is applied to the training set in two steps.

In the first step, the candidate attributes that are not significantly informative are screened out in a pre-selection process. Theoretically, any attribute that is non-informative, such as white noise, should have zero mutual information with the target variable. However, \cite{Steuer2002} noted that the estimation of mutual information can be affected by systematic errors resulting from the finite-size issue. Hence, a non-informative attribute might not have zero mutual information. So instead of using zero as a baseline, for two variables $X$ and $Y$, we create 1000 pairs of constraint realizations by randomly permutating their original data, and use the average of the mutual information of all the pairs as a baseline. Then, any candidate attribute with a smaller mutual information value than the baseline is deemed not significantly informative and is thus deleted.

In the second step, an mRMR algorithm is used in combination with a forward selection process to select a set of highly informative but minimally redundant features, as described in section \ref{mRMR}. The candidate attribute with the highest information level is used as the starting point. A pre-test procedure is conducted to determine the optimal number of features. Since the F-measure is more efficient to evaluate specifically the jump instance prediction accuracy, it is used as the primary evaluation metric in the feature selection procedure. We randomly select 70\% of the training instances and use them to construct models with various numbers of features. Then, the optimal number of features is determined by comparing the F-measure of these models on the remaining 30\% of the training instances.

\subsection{Jump prediction}
\subsubsection{Model training}

Support vector machines (SVMs), artificial neural networks (ANNs), random forest (RFs) and K-nearest neighbours (KNNs) are four state-of-the-art machine learning algorithms to handle large numbers of input features and their complex nonlinear relationships with the target variable. In our study, all four algorithms are utilized and compared to develop the optimal jump prediction model.

In an SVM classifier, the most commonly used Gaussian kernel is adopted, where the trade-off parameter $c$ and the kernel scale parameter $g$ are important parameters to be optimized.

In an ANN classifier, a feed-forward neural network consisting an input lay, a hidden layer and an output layer is used. A sigmoid function and a soft max function are the activation functions respectively in the hidden layer and the output layer. Scaled conjugate gradient descent method is utilized to compute the optimal weights in each epoch where cross-entropy is adopted as the loss function. The number of neurons $nn$ in the hidden layer needs to be further optimized.

For RF and KNN classifiers, the number of trees $tr$ and the number of neighbours $ngb$ need to be optimized respectively.

The value range of the important parameters in each algorithm are listed in Table \ref{tab:algoParam}.
\begin{table}[htbp]
\begin{center}
\caption{Important parameters and their value ranges of machine learning algorithms}\label{tab:algoParam}
\begin{tabular}{lllllllllllllll}
\hline\noalign{\smallskip}
Algorithm             &Parameter           &Value range       \\
\noalign{\smallskip}\hline\noalign{\smallskip}

\multirow{2}{*}{SVM}             &$c$           &0.01, 0.1, 1, 10, 100\\
                        &$g$       &0.01, 0.1, 1, 10, 100\\
ANN             &$nn$    &5, 10, 15, 20, 30, 40\\
RF              &$tr$    &10, 30, 50, 100, 200\\
KNN             &$ngb$   &5, 10, 30, 50, 100\\
\noalign{\smallskip}\hline
\end{tabular}
\end{center}
\end{table}

Optimization of these parameters is through exhaustive searching. That is, for each value (or each pair of values) of the parameter(s), we choose the optimal number of features as described in section \ref{ftselect}. Then the optimal value(s) of parameter(s) are determined when the algorithm gives the highest prediction performance with the corresponding optimal number of features.

\subsubsection{Model validation}
Assume that the numbers of the upward jumping, downward jumping and non-jumping instances in a testing dataset are $\tilde{M}, \tilde{N}$ and $\tilde{L}$. Since $\tilde{L}$ should still be hundreds times larger than $\tilde{M}$ or $\tilde{N}$ as in the training set, the prediction accuracy on the non-jumping class will dominate that on the whole testing set. Thus to alleviate this problem, we evaluate a model performance on all the jumping instances, but only on subsets of non-jumping ones. That is, we randomly sample 50 subsets from the non-jumping class to construct 50 replicate testing sets $repTe_j, j = 1, 2, \cdots, 50$. The size of the subset is taken $\tilde{M}+\tilde{N}$ in the binary classification scenario, and $\max(\tilde{M}, \tilde{N})$ in the trinary classification scenario, so that the size of each of the $repTe_j (j = 1, 2, \cdots, 50)$ should be $2(\tilde{M}+\tilde{N})$ and $\tilde{M}+\tilde{N}+\max(\tilde{M}, \tilde{N})$ respectively.

\begin{figure}[htbp]
\begin{center}
\includegraphics[scale=0.42]{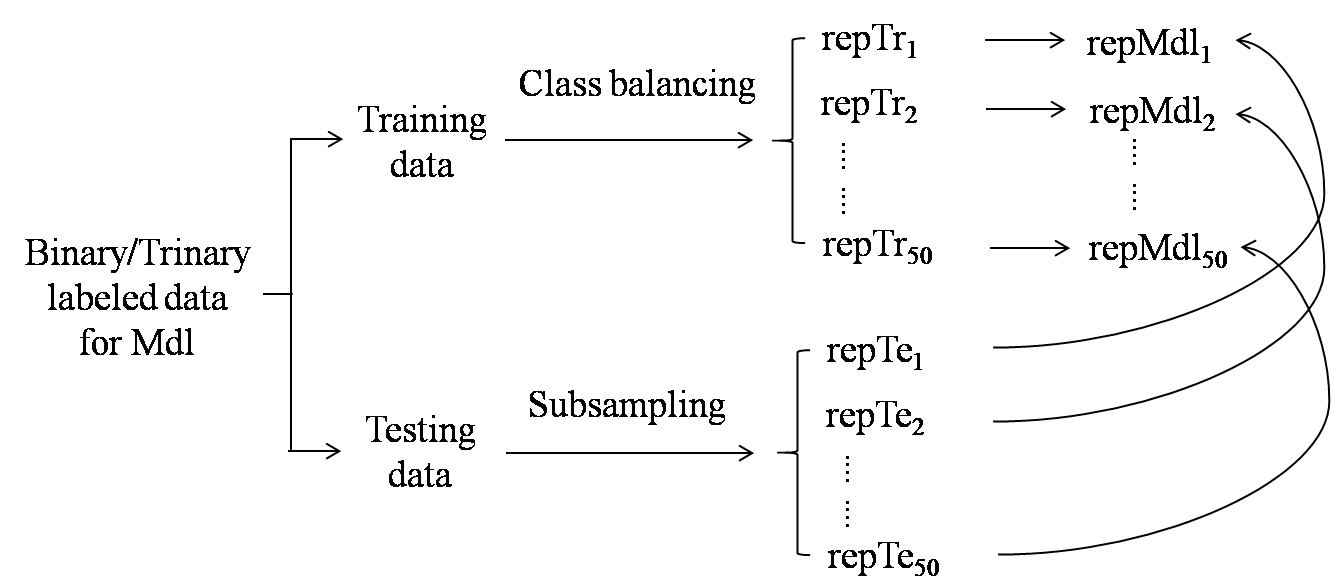}
\caption{Terminology of model building in our approach}\label{Fig:replicate}
\end{center}
\end{figure}

Remember that for a certain type of model $Mdl$, where $Mdl$ can be a $CompMdl$ or a $IndMdl(i)$, $(i = 1, 2, \cdots, n)$, we construct 50 replicate training sets $repTr_j, j = 1, 2, \cdots, 50$ after class balancing in section \ref{balance}. Then as shown in Figure \ref{Fig:replicate}, for each of the replicate model $repMdl_j$ of $Mdl$ trained on $repTr_j$, one of the replicate testing sets $repTe_j$ without repeating is used to evaluate its accuracy and F-measure defined in section \ref{eval}. Then the expected performance as well as its variation of $Mdl$ can be computed from the 50 performance samples.

\section{Empirical results}\label{empirical}
\subsection{Data}\label{data}
Our study is based on the level-2 transaction data of all the main-board and second-board stocks on the Shenzhen Stock Exchange of China from January 2014 to August 2017 (a total of 896 trading days). The level-2 data, which are updated every 3 seconds, consist of the best ten quotes, the current transaction price, the current transaction volume, the cumulative number of trades and volume from the last record to the current record. The data are downloaded from Wind database. A total of 2019 stocks are included, but to ensure the adequacy of the research data, only the 1271 stocks that are traded on more than 90\% of the 896 days are studied. The dataset is then divided into two parts: the data from January 2014 to December 2015 are used to train the prediction models and the rest are used for model validation.

The Shenzhen Stock Exchange is open from 9:30 am to 11:30 am and 1:00 pm to 3:00 pm, with a total of 4 hours in a trading day. In our experiment, as described in section \ref{design}, each trading day is divided into 48 five-minute intervals.

\subsection{Detected jumps}
\begin{figure}
\begin{center}
\includegraphics[scale=0.23]{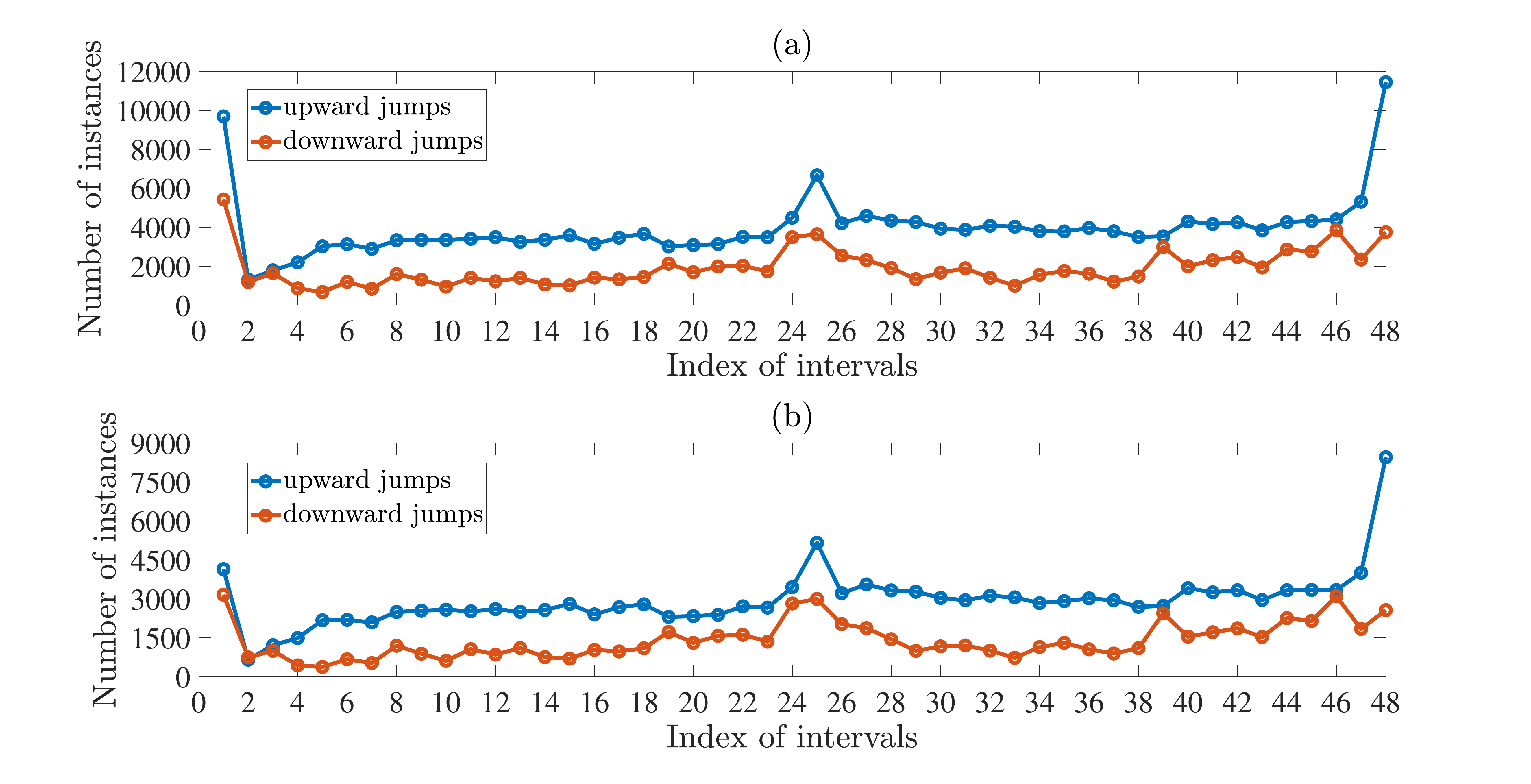}
\caption{(a) Number of detected jumps in each interval (b) Number of jumps in each interval after reduction}\label{jumpfig}
\end{center}
\end{figure}

Intraday jumps are detected via the technique described in section \ref{detection}. Figure \ref{jumpfig}(a) displays the number of detected jumps for the whole dataset in each interval of a trading day.

The numbers of upward and downward jumps exhibit a U-shaped trend and an L-shaped trend, respectively, within a day. The number of jumps peaks in the first five minutes after opening and falls sharply in the second five minutes. Then, the values remain relatively stable until the close of the morning session. This feature indicates that the accumulation of overnight information has a significant impact on stock jumps. Relatively smaller peaks are also observed after the noon break due to the accumulation of information during the break. The number of upward jumps increases rapidly and sharply to its highest value in the last 5 minute of a trading day because most investors are not well-informed and have a wait-and-see attitude until the last minute. The number of downward jumps fluctuates more substantially during the last hour, yet no peak is observed. This result indicates that investors are not eager to sell after they receive bad news. Alternatively, they constantly search for suitable shorting time to the last minute.

These phenomena verify the time-varying characteristic of stock jumps in the Chinese market. Thus, to explore the possibility of time-varying predictability of stock jumps, our approach attempts to construct prediction models every five minutes in addition to a comprehensive model for all the time intervals.

We impose four additional conditions before including jumping instances in the training and testing data. First, there is a 10\% price change limit rule in the Chinese stock market, and no transactions can be made after the price limit is reached. Therefore, among all sequential limit-ups or limit-downs, only the first, if is recognized as a jump, is considered. Second, because of the implementation of a circuit breaker mechanism on the Chinese stock market from 2016/01/04 to 2016/01/07, the whole stock market halted several times on 2016/01/04 and 2016/01/07; thus, jumps that occurred between 2016/01/04 and 2016/01/08 are deleted. Third, a jump must occur at least an hour after stock suspension or dividend distribution; this criterion ensures the consistency of the information window. Fourth, since a 60-day window is used to normalize the liquidity measures and technical indicators, as illustrated in section \ref{lm} and \ref{ti}, jumps occur in the first 60 trading days of each stock are not considered.

After applying these conditions, approximately 25\% of the jumps are deleted in all intervals except for the first few ones, in which much more than 25\% of the jumps are removed, as shown in Figure \ref{jumpfig}(b). This difference is because in Chinese stock market, most of the sequential limit-ups or limit-downs, tend to occur at the beginning of a trading day; though recognized as jumps, they should be deleted except for the first one in each sequence, due to the insufficiency of ex ante market information.

\subsection{Predictive power of candidate attributes}

After jump detection and attribute computation, an instance, whether from the training period or the testing period, is represented by a vector of 184 attribute values, of which 130 are calculated from liquidity measures and 54 are from technical indicators. For the individual model building, the training and testing set of instances are further split into 48 sets corresponding to the 48 five-minute intervals.

To reduce the effect of unbalanced data, a random undersampling procedure is performed to balance the two classes in the binary classification problem, while a SMOTE technique is performed in addition to balance the three classes in the trinary classification problem, as described in section \ref{balance}. Fifty times of such balancing procedures are repeated to construct fifty replicates of training sets with balanced data.

\begin{figure}[htbp]
\begin{center}
\includegraphics[scale=0.23]{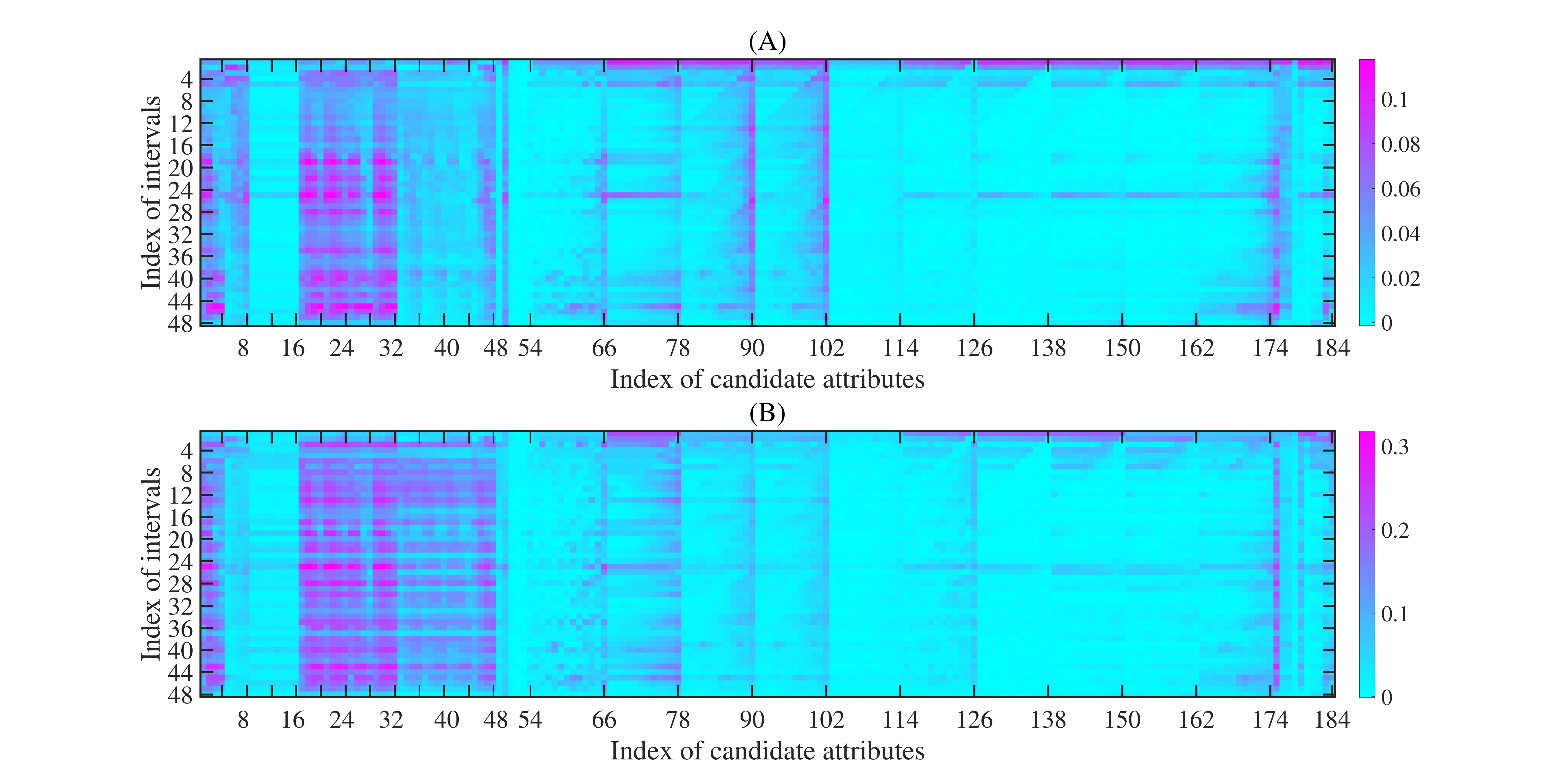}\\
\caption{Heatmap of excess mutual information between stock jumps with candidate attributes in all 5-minute intervals regardless of the jump directions (A) or regard of the jump directions (B). The excess mutual information is computed by subtracting the baseline value from the mutual information.}\label{fig:MIfig}
\end{center}
\end{figure}

Then, the two-step feature selection procedure elaborated in section \ref{ftselect} is applied to each of the training replicates to select a set of highly informative but minimally redundant features from the 184 candidate attributes based on mutual information. Mutual information is computed by the adaptive partition technique where the number of bins is 100 or 30 for all the training data or for the training data that is divided into 48 intervals. In the first step, an attribute is deleted if the informative level is below the baseline, where the baseline is computed from permuted original data. Figure \ref{fig:MIfig} exhibits the excess informative level averaged over 50 training replicates of all the candidate attributes in contrast to the baseline regard or regardless of the jump directions, where the indices of these attributes are given in Table \ref{attridx}.

\begin{table}
\centering
\caption{Index of candidate attributes}\label{attridx}
\begin{tabular}{lllll}
\hline\noalign{\smallskip}
Categories        &Candidate attributes              &Index            \\
\noalign{\smallskip}\hline\noalign{\smallskip}
Technical indicators       &$PROC$             &1-4\\
with 5,10,20,30 lagged intervals   &$VROC$             &5-8\\
                            &$MA$              &9-12\\
                            &$EMA$             &13-16\\
                            &$BIAS$            &17-20\\
                            &$EBIAS$           &21-24\\
                            &$OSCP$            &25-28\\
                            &$EOSCP$           &29-32\\
                            &$fK$              &33-36\\
                            &$fD$              &37-40\\
                            &$sD$              &41-44\\
                            &$CCI$             &45-48\\
                            &$ADO$             &49\\
                            &$TR$              &50\\
                            &$PVT$              &51\\
                            &$OBV$              &52\\
                            &$NVI$              &53\\
                            &$PVI$              &54\\
\noalign{\smallskip}\hline\noalign{\smallskip}
Liquidity measures          &$r$                &55-66\\
of prior 12, 11, $\cdots$, 2,1 intervals     &$R$                &67-78\\
                            &$k$                &79-90\\
                            &$v$                &91-102\\
                            &$s$                &103-114\\
                            &$oi$               &115-126\\
                            &$di$               &127-138\\
                            &$qs$               &139-150\\
                            &$es$               &151-162\\
                            &$rv$               &163-174\\
                            &average of $r$     &175\\
                            &average of $R$     &176\\
                            &average of $k$     &177\\
                            &average of $v$     &178\\
                            &average of $s$     &179\\
                            &average of $oi$     &180\\
                            &average of $di$     &181\\
                            &average of $qs$     &182\\
                            &average of $es$     &183\\
                            &average of $rv$     &184\\
\noalign{\smallskip}\hline
\end{tabular}
\end{table}

In both scenarios where the jump direction is considered or not, we observe that the technical indicators are in general more informative than the liquidity measures, except in the first few intervals. PROC, BIAS, EBIAS, OSCP and EOSCP (indices 1-4, 17-20, 21-24, 25-28 and 29-32) are the most informative technical indicators. Indicators such as fK, fD, sD, CCI and TR (indices 33-36, 37-40, 41-44, 45-48, 50) also show high information levels, especially when we consider the jump directions.

Among the liquidity measures, the most informative one is the average value of returns over the prior 12 intervals(index 175) in both scenarios. As to the measures computed in individual intervals, the accumulative return (indices 67-78), the number of trades (indices 79-90) and the trading volume (indices 91-102) are more informative than others. Furthermore, the figures show that the information level of a liquidity measure improves with increasing interval index, which is reasonable since a higher index is related to a closer interval to the current instance and thus leads to higher predictive power.

\subsection{Feature selection and parameter setting}
In the second step of feature selection using the mRMR algorithm, the optimal number of features is chosen among the values of $\{5, 10, 15, 20, \cdots, maxno\}$, where $maxno$ denotes the number of retained features after the significantly non-informative ones are deleted. Note that $maxno$ is different in various scenarios. For the comprehensive model building, the feature screening result actually shows that there is no redundant features, so the total number of retained features is 184; while for the individual model building in each interval, the $maxno$ is usually smaller than 184, indicating that several features are not very informative in certain intervals. The best number of features leading to the best prediction performance, especially the best F-measure, is chosen through examining every possible list of mRMR selected features. The prediction performance is computed by building the prediction model on a randomly selected 70\% of the training data and testing it on the remaining 30\% of the data. Then the optimal values of parameters are set with the ones that can achieve the highest prediction performance with the corresponding optimal number of features.

Tables \ref{tab:NoFT} gives the optimal number of features, the algorithm parameter setting and the corresponding prediction performance over 50 replicate training sets for both the comprehensive and individual models regard or regardless of the jump direction. In a comprehensive model $CompMdl$, the size of each replicate training set is usually more than 200,000. In an individual model $IndMdl(i)$, the size of each replicate training set varies from around 1,000 to 10,000. It can be seen from Table \ref{tab:NoFT} that though all the algorithms went through a feature selection procedure, they all tend to use the maximal number of features after examining all lists of selected features. This indicates that all the used algorithms are capable of exploiting data patterns from large dimensional inputs, especially when the training sets are as large as in our study.

\begin{table}[htbp]\small
\begin{center}
\caption{Number of selected features and model performance on the training sets}\label{tab:NoFT}
\begin{tabular}{clccccccc}
\hline\noalign{\smallskip}
                        & &SVM           &ANN        &RF         &KNN               \\
\noalign{\smallskip}\hline\noalign{\smallskip}
\multicolumn{3}{l}{\underline{jump/no jump}}\\\noalign{\smallskip}
\multirow{4}{*}{ $CompMdl$ }  &Param.     &$(c,g):(1,10)$    &$nn:10$    &$tr:100$      &$ngb:30$        \\\noalign{\smallskip}
                        &$N_{FT}$     &184          &184             &184            &184                 \\\noalign{\smallskip}
                        &$fm$ (\%)    &66.6            &64.4             &68.2               &62.7            \\\noalign{\smallskip}
                        &$acc$ (\%)   &69.4            &66.7             &69.9               &64.5            \\\noalign{\smallskip}
\\
 \multirow{4}{*}{$IndMdl$}&Param.             &$(c,g):(1,10)$     &$nn:10$    &$tr:100$  &$ngb:30$            \\\noalign{\smallskip}
                        &$N_{FT}$           &$maxno$            &$maxno$        &$maxno$     &$maxno$        \\\noalign{\smallskip}
                        &$\overline{fm}$ (\%)  &67.6           &64.7       &69.1       &64.0             \\\noalign{\smallskip}
                        &$\overline{acc}$ (\%)   &69.5         &66.9       &70.7       &66.0            \\\noalign{\smallskip}
\noalign{\smallskip}\hline\noalign{\smallskip}
\multicolumn{3}{l}{\underline{upward/downward/no jump}}\\\noalign{\smallskip}
\multirow{5}{*}{$CompMdl$}  &Param.                 &(c,g):(1,10)       &$nn:30$     &$tr:100$  &$ngb:30$          \\
                        &$N_{FT}$                   &184     &184       &184        &184            \\\noalign{\smallskip}
                        &$\overline{fm}^+$ (\%)    &61.4        &59.3                 &62.8            &53.5               \\\noalign{\smallskip}
                        &$\overline{fm}^-$ (\%)    &74.3        &69.9                 &76.5            &64.5               \\\noalign{\smallskip}
                        &$\overline{acc}$ (\%)   &66.1          &62.6                  &68.0            &56.9               \\\noalign{\smallskip}
                        \\
\multirow{5}{*}{$IndMdl$}&Param.                  &$(c,g):(1,10)$     &$nn:15$        &$tr:100$       &$ngb:30$         \\\noalign{\smallskip}
                       &$N_{FT}$                &$maxno$            &$maxno$        &$maxno$        &$maxno$           \\\noalign{\smallskip}
                       &$\overline{fm}^+$ (\%)  &63.3               &60.1           &65.5           &53.3               \\\noalign{\smallskip}
                       &$\overline{fm}^-$ (\%)  &77.8               &70.3           &80.0           &70.0              \\\noalign{\smallskip}
                       &$\overline{acc}$ (\%)   &68.0               &61.9           &70.7           &59.6               \\\noalign{\smallskip}
\noalign{\smallskip}\hline
\end{tabular}
\end{center}{For a $CompMdl$, the optimal number of features is chosen based on the average performance over 50 replicate training sets. For $IndMdl(i)$, the optimal number of features is chosen based on the average performance over all $IndMdl(i) (i = 1,2, \cdots, 48)$ and over 50 replicate training sets.}
\end{table}

\subsection{Model performance}
Table \ref{tab:perfall} summarizes the performance of the four machine learners in building comprehensive and individual models for jump prediction regard or regardless of the jump direction. It can be seen that in general predicting whether a stock jump is arriving is easier than predicting the its direction. Clearly, for all the prediction problems, RF seems to be the most promising model, with the highest average prediction performance and the smallest performance variation. The performance of SVM is slightly higher than ANN, and that of KNN ranks the lowest.

In addition, we find that, comprehensive models always outperform individual models. This might because the data used to construct a comprehensive model is much larger than that to each individual model; besides, the algorithm parameters are not optimized to adapt each individual model. The performance of individual models might be further improved through adaptively optimizing parameters in each interval.

\begin{table}[htbp]\small
\begin{center}
\caption{Prediction performance of machine learning algorithms with comprehensive and individual models on the testing sets (\%)}\label{tab:perfall}
\begin{tabular}{clccccccccccc}
\hline\noalign{\smallskip}
                        &                       &SVM          &ANN                  &RF                     &KNN              \\
\noalign{\smallskip}\hline\noalign{\smallskip}
\multicolumn{3}{l}{\underline{jump/no jump}}\\\noalign{\smallskip}
\multirow{2}{*}{$CompMdl$}  &$fm$ (\%)              &$60.4\pm0.3$           &$60.1\pm1.2$           &$\bf{63.1\pm0.2}$         &$60.5\pm0.2$         \\\noalign{\smallskip}
                        &$acc$ (\%)              &$67.0\pm0.2$          &$65.2\pm1.2$           &$\bf{67.7\pm0.1}$         &$63.3\pm0.2$         \\\noalign{\smallskip}
\multirow{2}{*}{$IndMdl$}  &$\overline{fm}$ (\%)    &$57.9\pm0.3$       		&$55.5\pm0.9$            &$\bf{60.7\pm0.2}$ 	 		&$58.6\pm0.3$           \\\noalign{\smallskip}
                        &$\overline{acc}$ (\%)   &$65.4\pm0.2$       		&$63.0\pm0.7$       	&$\bf{66.6\pm0.1}$	      	&$62.9\pm0.2$            \\\noalign{\smallskip}
\noalign{\smallskip}\multicolumn{3}{l}{\underline{upward/downward/no jump}}\\
\multirow{3}{*}{$CompMdl$}  &$fm^+$ (\%)           &$56.3\pm0.3$                  &$55.2\pm0.6$                  &$\bf{57.6\pm0.2}$      &$52.9\pm0.3$                      \\\noalign{\smallskip}
                        &$fm^-$ (\%)           &$56.3\pm0.5$                  &$56.7\pm0.7$                  &$\bf{58.2\pm0.1}$      &$52.0\pm0.2$                        \\\noalign{\smallskip}
                        &$acc$ (\%)            &$59.4\pm0.2$                  &$58.5\pm0.3$                  &$\bf{60.0\pm0.2}$      &$52.8\pm0.2$                        \\\noalign{\smallskip}
\multirow{3}{*}{$IndMdl$}  &$\overline{fm}^+$ (\%) &$56.7\pm0.3$   &$55.5\pm0.5$                  &$\bf{57.8\pm0.2}$       &$53.6\pm0.3$               \\\noalign{\smallskip}
                        &$\overline{fm}^-$ (\%)  &$45.5\pm0.6$  &$45.0\pm0.7$                   &$\bf{49.0\pm0.5}$       &$44.2\pm0.3$           \\\noalign{\smallskip}
                        &$\overline{acc}$ (\%)   &$57.8\pm0.3$  &$54.7\pm0.5$                   &$\bf{59.3\pm0.2}$       &$51.8\pm0.3$          \\\noalign{\smallskip}
\noalign{\smallskip}\hline
\end{tabular}
\end{center}{The results are computed through averaging over 50 replicate models. The best performance among the four algorithms at a 0.05 significance level is boldfaced}
\end{table}

\begin{figure}[htbp]
\begin{center}
\includegraphics[scale=0.22]{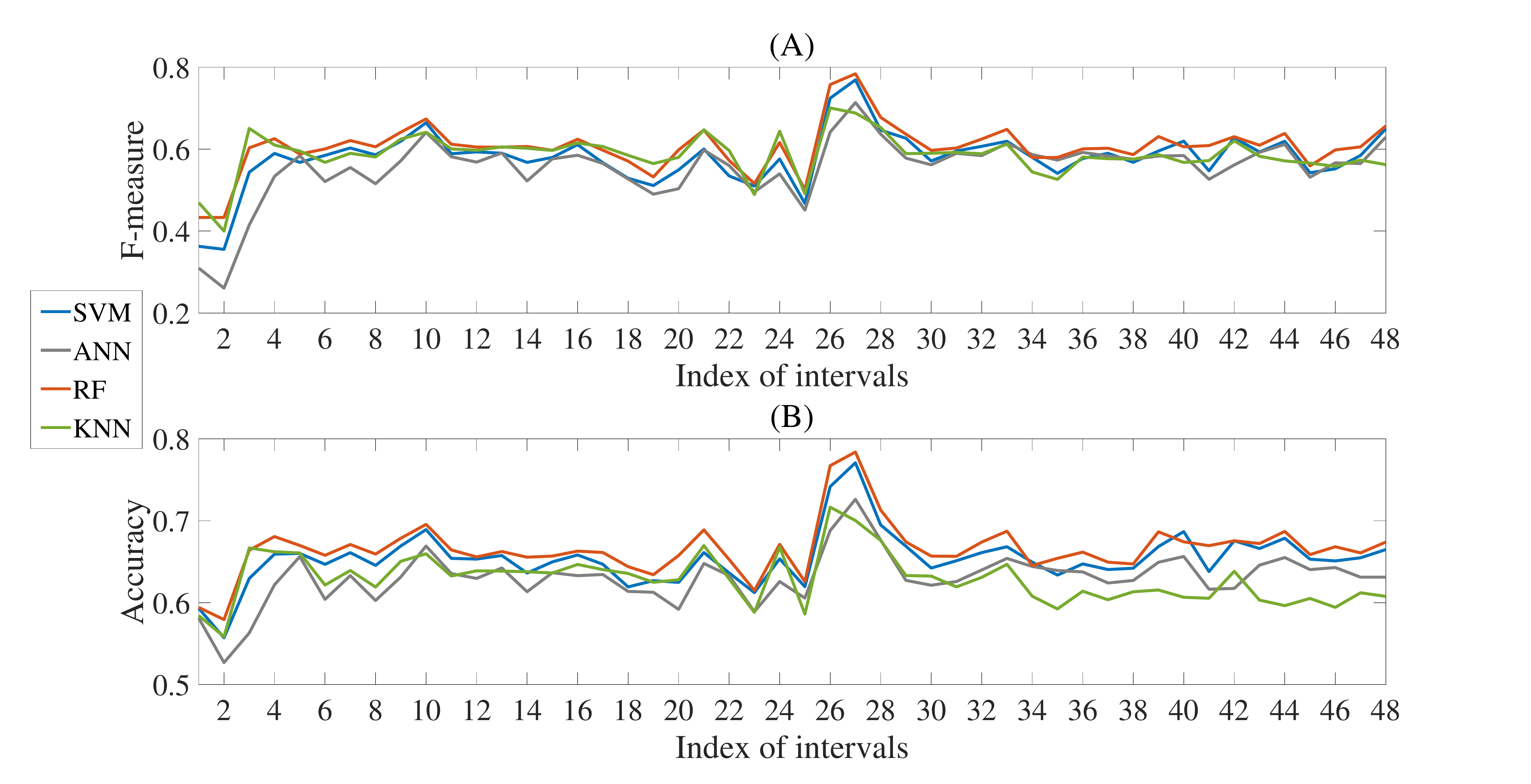}\\
\caption{F-measure (A) and accuracy (B) of five machine learning algorithms within each interval in predicting stock jumps regardless of the jump directions. }\label{perf2class}
\end{center}
\end{figure}

\begin{figure}[htbp]
\begin{center}
\includegraphics[scale=0.22]{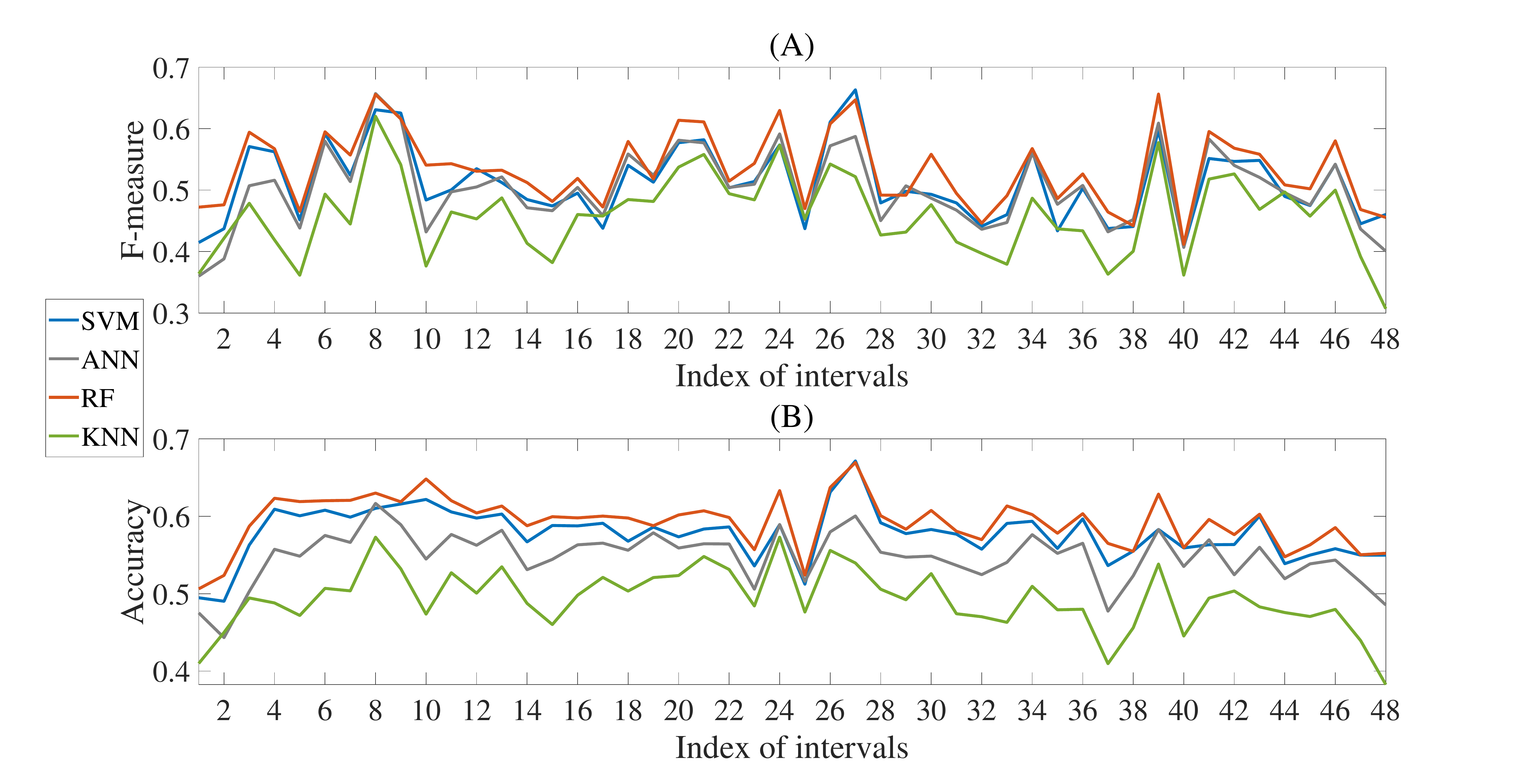}\\
\caption{F-measure (A) and accuracy (B) of five machine learning algorithms within each interval in predicting stock jumps regard of the jump directions. F-measure is plotted on its average values of both upward and downward jumps. }\label{perf3class}
\end{center}
\end{figure}

Figure \ref{perf2class} and \ref{perf3class} show the F-measure and accuracy of the individual models within each interval in the binary and trinary classification problems. It is still obvious that RF can achieve higher F-measures and accuracies than other algorithms in most of the intervals.

To demonstrate the efficiency of using comprehensively two types of candidate attributes, Table \ref{tab:perfrf} compares the performance of the best learner, RF, on the sets of liquidity measures, technical indicators and both of them. Obviously, using both types of attributes simultaneously is superior to using only one of them in any scenario. This indicates that there is difference between the information of technical indicators and liquidity measures for jump prediction. It is also notable that comparing between liquidity measures and technical indicators, the latter is more capable to predict both the arrival of stock jumps or the jump directions, which is in line with the informative levels exhibited in Figure \ref{fig:MIfig}.

\begin{table}[htbp]\small
\begin{center}
\caption{RF performance on difference sets of candidate attributes (\%)}\label{tab:perfrf}
\begin{tabular}{llccccccccccc}
\hline\noalign{\smallskip}
                        &                       &Liquidity          &Technical                  &Both\\
\noalign{\smallskip}\hline\noalign{\smallskip}
\multicolumn{3}{l}{\underline{jump/no jump}}\\\noalign{\smallskip}
\multirow{2}{*}{$CompMdl$}  &$fm$ (\%)              &$61.2\pm0.2$	          &$62.4\pm0.2$      &$\bf{63.1\pm0.2}$  \\\noalign{\smallskip}
                        &$acc$ (\%)              &$66.7\pm0.2$            &$66.8\pm0.1$           &$\bf{67.7\pm0.1}$ \\\noalign{\smallskip}
\multirow{2}{*}{$IndMdl$}  &$\overline{fm}$ (\%)    &$58.5\pm0.2$           &$60.5\pm0.2$           &$60.7\pm0.2$\\\noalign{\smallskip}
                        &$\overline{acc}$ (\%)   &$65.1\pm0.2$            &$65.8\pm0.1$            &$\bf{66.6\pm0.1}$\\\noalign{\smallskip}
\noalign{\smallskip}\multicolumn{3}{l}{\underline{upward/downward/no jump}}\\
\multirow{3}{*}{$CompMdl$}  &$fm^+$ (\%)           &$55.7\pm0.2$                &$56.7\pm0.3$           &$\bf{57.6\pm0.2}$\\\noalign{\smallskip}
                        &$fm^-$ (\%)           &$56.1\pm0.2$                &$57.0\pm0.3$           &$\bf{58.2\pm0.1}$ \\\noalign{\smallskip}
                        &$acc$ (\%)            & $58.8\pm0.2$               &$59.0\pm0.2$           &$\bf{60.0\pm0.2}$\\\noalign{\smallskip}
\multirow{3}{*}{$IndMdl$}  &$\overline{fm}^+$ (\%) &$55.0\pm0.3$              &$56.9\pm0.3$           &$\bf{57.8\pm0.2}$\\\noalign{\smallskip}
                        &$\overline{fm}^-$ (\%)  &$42.7\pm0.6$              &$48.9\pm0.5$           &$49.0\pm0.5$\\\noalign{\smallskip}
                        &$\overline{acc}$ (\%)   &$56.8\pm0.2$              &$58.2\pm0.2$           &$\bf{59.3\pm0.2}$\\\noalign{\smallskip}
\noalign{\smallskip}\hline
\end{tabular}
\end{center}{The results are computed through averaging over 50 replicate models. The best performance among the three types of attributes at a 0.05 significance level is boldfaced.}
\end{table}

\subsection{Feature importance analysis}
Random forest is an ensemble of trees constructed from bootstrap instances of the training set. As it randomly selects features in each of the tree induction processes, it can be used to rank the importance of features in a natural way. To measure the importance of the $j$th feature, the values of the $j$th feature are perturbed and the averaged difference in the out-of-bag error before and after the permutation over all trees is computed as the importance score for the $j$th feature. Features with large scores (normalized by the standard deviation) are ranked as more important ones.

\begin{figure}[htbp]
\begin{center}
\includegraphics[scale=0.22]{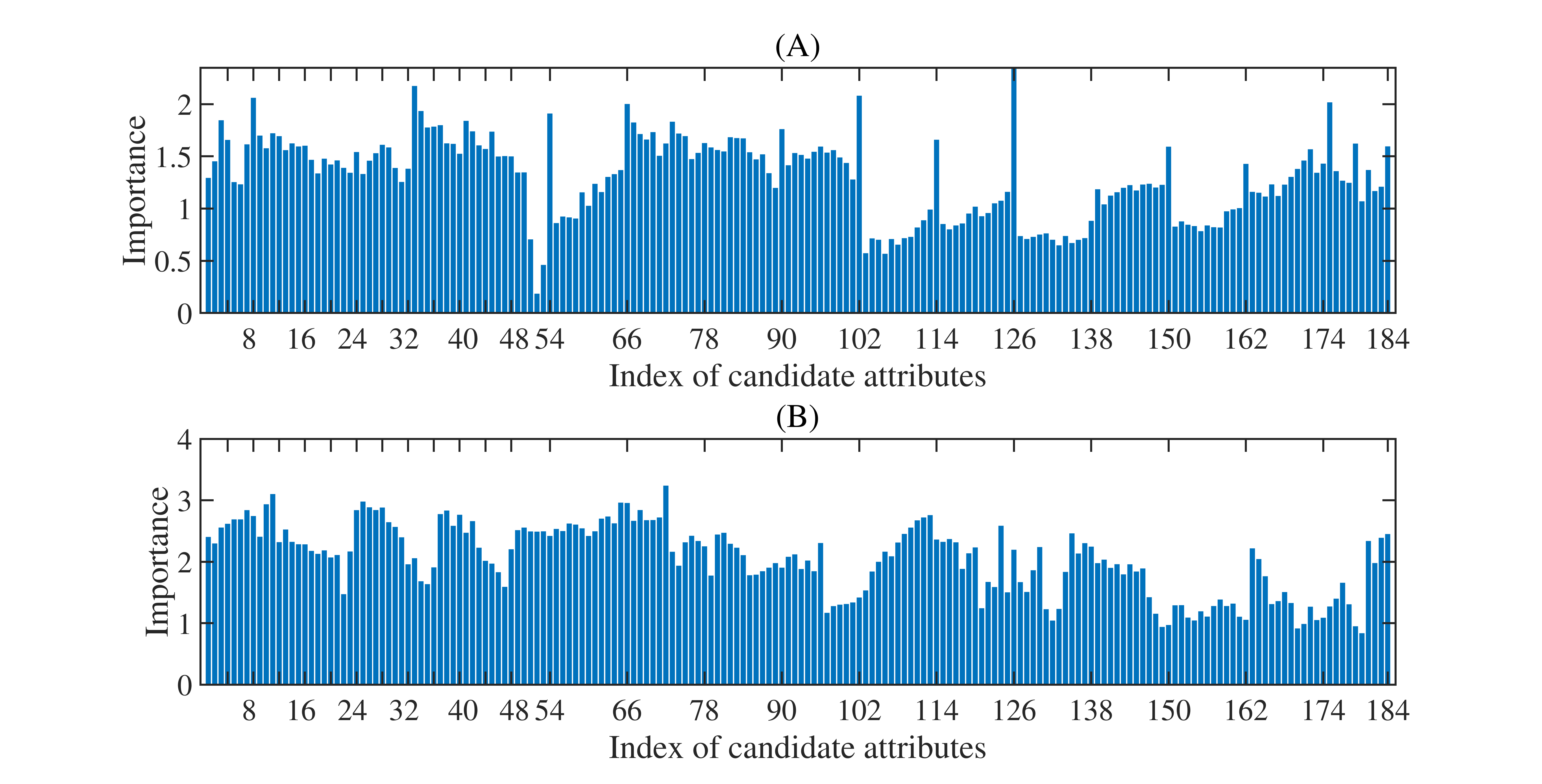}\\
\caption{Feature importance analysis by random forest in predicting stock jumps regardless of the jump directions (A) or regard of the jump directions(B).}\label{fig:ftimp}
\end{center}
\end{figure}

We examine the importance of our 184 attributes for jump prediction with the random forest. That is, we adopt all the 184 attributes as input features of the training data, and construct a comprehensive model due to its superior performance over individual models. Taking into account the subsampling variation, 50 replicates of training sets are used to compute 50 times of feature importance. Figure \ref{fig:ftimp} shows the plot of the feature importance in both the binary and trinary classification scenarios. Generally, there is no significant difference between the importance of liquidity measures and technical indicators. Most of the features have comparable contribution to the prediction, even those with low informative level as shown in Figure \ref{fig:MIfig}. This verifies the fact that the features we construct are effective for stock jump prediction. Because individual features are always of low predictive power, machine learners tend to use most of the features to reach the optimal prediction performance.

\subsection{Robustness test}\label{robust}

For a robustness check, we repeat the same prediction procedure based on the jump instances detected at a significant level of 1\%. For better comparison, we use the same parameters as in Table \ref{tab:algoParam} for all the algorithms. The prediction results of comprehensive models and individual models in both binary and trinary classification scenarios are given in Table \ref{tab:perfallrobust}. Clearly, all these results are highly consistent with those obtained when the significance level of jump detection is 5\%, which supports the effectiveness of our jump prediction approach.

\begin{table}[htbp]\small
\begin{center}
\caption{Robustness test: Prediction performance of machine learning algorithms with comprehensive and individual models on the testing sets (\%)}\label{tab:perfallrobust}
\begin{tabular}{llccccccccccc}
\hline\noalign{\smallskip}
                        &                       &SVM          &ANN                  &RF                     &KNN              \\
\noalign{\smallskip}\hline\noalign{\smallskip}
\multicolumn{3}{l}{\underline{jump/nojump}}\\\noalign{\smallskip}
\multirow{2}{*}{$CompMdl$}  &$fm$ (\%)              &$61.3\pm0.3$     &$61.0\pm1.2$         &$\bf{63.7\pm0.2}$       &$61.1\pm0.2$             \\\noalign{\smallskip}
                        &$acc$ (\%)              &$67.6\pm0.2$    &$65.8\pm1.1$         &$\bf{68.2\pm0.1}$       &$63.9\pm0.2$           \\\noalign{\smallskip}
\multirow{2}{*}{$IndMdl$}  &$\overline{fm}$ (\%)    &$58.4\pm0.3$   &$56.0\pm1.0$         &$\bf{61.0\pm0.2}$       &$58.9\pm0.1$     		\\\noalign{\smallskip}
                        &$\overline{acc}$ (\%)   &$65.7\pm0.2$    &$63.4\pm0.4$    		&$\bf{66.8\pm0.1}$       &$63.3\pm0.1$        \\\noalign{\smallskip}
\noalign{\smallskip}\multicolumn{3}{l}{\underline{upward/downward/no jump}}\\
\multirow{3}{*}{$CompMdl$}  &$fm^+$ (\%)      &$57.3\pm0.2$         &$56.5\pm0.4$           &$\bf{58.6\pm0.2}$           &$53.7\pm0.3$                 \\\noalign{\smallskip}
                        &$fm^-$ (\%)      &$58.0\pm0.2$         &$58.8\pm0.7$           &$\bf{59.8\pm0.1}$           &$53.1\pm0.2$                  \\\noalign{\smallskip}
                        &$acc$ (\%)       &$60.6\pm0.2$         &$59.6\pm0.3$           &$\bf{61.2\pm0.2}$           &$53.7\pm0.2$                  \\\noalign{\smallskip}
\multirow{3}{*}{$IndMdl$}  &$\overline{fm}^+$ (\%) &$57.9\pm0.3$         &$56.6\pm0.5$           &$\bf{58.8\pm0.3}$           &$54.7\pm0.3$    \\\noalign{\smallskip}
                        &$\overline{fm}^-$ (\%) &$45.8\pm0.4$         &$44.8\pm0.9$           &$\bf{49.5\pm0.5}$           &$44.3\pm0.3$     \\\noalign{\smallskip}
                        &$\overline{acc}$ (\%)  &$58.7\pm0.3$         &$55.5\pm0.5$           &$\bf{60.3\pm0.2}$           &$52.9\pm0.3$     \\\noalign{\smallskip}
\noalign{\smallskip}\hline
\end{tabular}
\end{center}{The results are computed through averaging over 50 replicate models. The best performance among the four algorithms at a 0.05 significance level is boldfaced.}
\end{table}

\section{Comparison to existing literature}
To the best of our knowledge, few work has been published about the prediction of stock jumps and only \citet{Makinen2019} reports the prediction performance of stock jumps using machine learning algorithms and high-frequency data information. We thus compare our result with their work. \citet{Makinen2019} focuses on five individual stock in NASDAQ over 360 days, and a convolutional LSTM attention model (CNN-LSTM-Attention), proposed in their paper, reports the best prediction performance, evaluated mainly with F-measure. Our optimal result is obtained using RF algorithm and comprehensive models.

\begin{figure}[htbp]
\begin{center}
\includegraphics[scale=0.22]{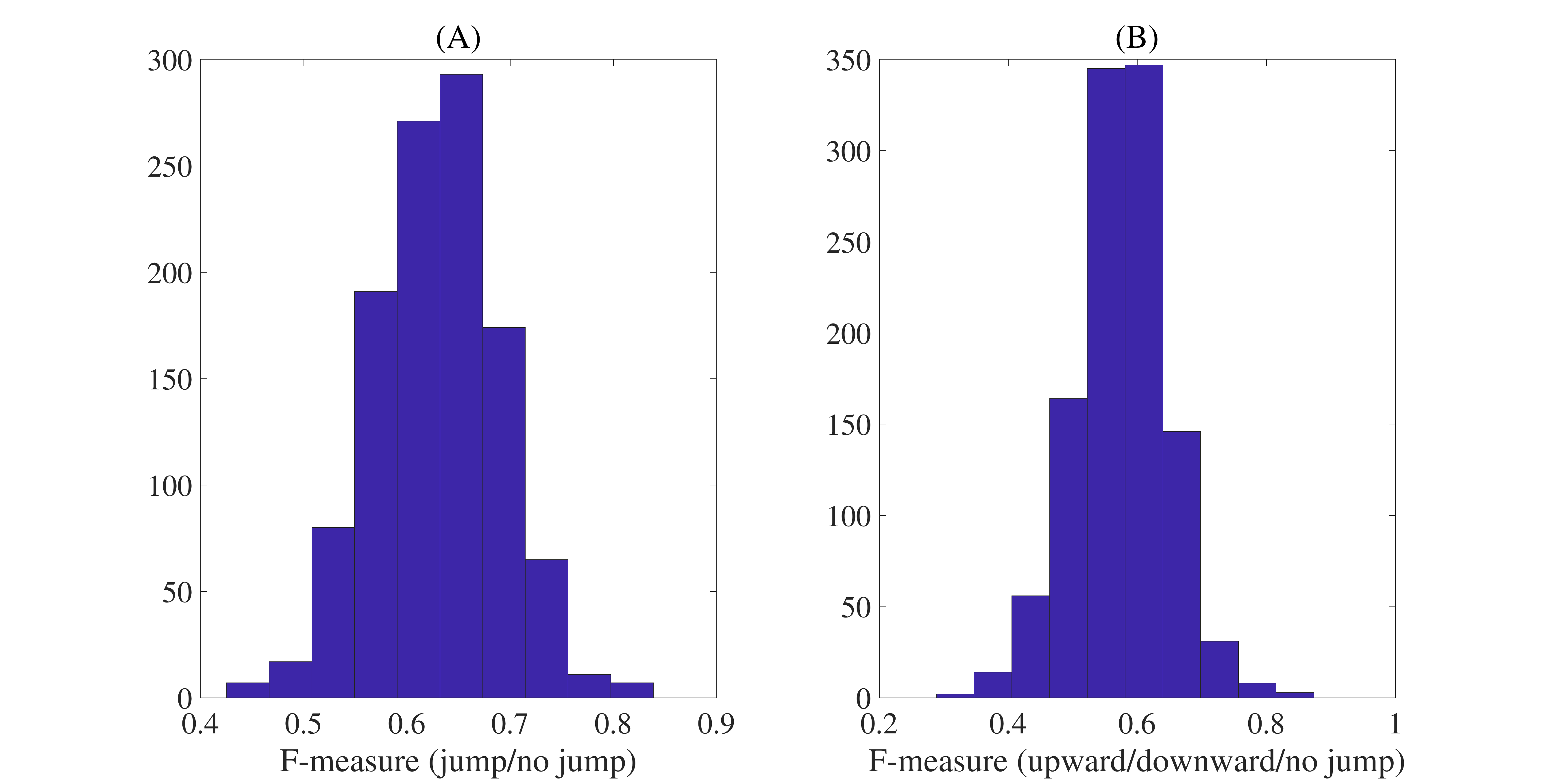}\\
\caption{F-measure of RF with comprehensive models in the binary classification (A) or the trinary classification problem. In both (A) and (B), F-measure is computed through averaging over 50 replicate models. In (B), F-measure is plotted on its average values of both upward and downward jumps. }\label{fig:fmhist}
\end{center}
\end{figure}

Regardless of the jump directions, \citet{Makinen2019} gives an average F-measure of 72\% over five well-known liquid stocks, GOOG, MSFT, AAPL, INTC and FB. Though our F-measure with RF and comprehensive model is only around 63\%, considering that our result is reported on 1271 stocks regardless of their liquidity, the performance of our model is still very competitive. One can find in Figure \ref{fig:fmhist}(A) that our study with RF can achieve F-measures of 72\% and above on more than 50 stocks among the 1271 ones (the actual number is 71). Moreover, our model is built universally for all stocks while \citet{Makinen2019}'s model is trained for individual stocks, the performance of which might raises underlying overfitting issues.

More interestingly, \citet{Makinen2019} reports an average F-measure of 49\% over the five stocks in predicting the jump directions, while our study can achieve a much higher F-measure around 59\%. Figure \ref{fig:fmhist}(B) shows the histogram of the F-measures evaluated by stocks. While RF model is simpler to the CNN-LSTM-Attention model, the significantly better results on a quite larger variety of stocks in our study imply that the liquidity measures and technical indicators constructed in this paper are more efficient than those attributes in  \citep{Makinen2019}.

\section{Conclusion}
In this paper, we propose a data-driven approach that combines liquidity measures and technical indicators to predict intraday stock jumps based on level-2 high-frequency data. Our proposed approach divides a complete trading day into sequential 5-minute intervals and forecasts the arrival of individual stock jumps every five minutes by exploiting the intraday information contained in 184 liquidity measures and technical indicators. This approach is then tested on the level-2 transaction data of all the main-board and second-board stocks on the Shenzhen Stock Exchange of China. Our results suggest that (i) the arrival of intraday stock jumps, as well as the direction of the jumps, can be predicted using technical indicators and liquidity measures; (ii) technical indicators have higher capacity to predict stock jumps than liquidity measures, but a combination of them is superior to either one in predicting stock jumps; (iii) based on the attributes provided in our study, RF outperforms SVM, ANN, and KNN. Importantly, our approach provides a standard framework to explore liquidity and technical information from level-2 data for the prediction of intraday jumps of individual stocks.

However, this study has its limitations and can be extended in the following ways. First, the parameters of individual models $IndMdl(i)(i = 1, 2, \cdots, n)$ are optimized universally for all $i$. To obtain higher performance, adaptive parameter optimization for each $IndMdl(i)$ can be implemented. Second, SMOTE technique is adopted to balance the upward and downward jumping classes. To improve the prediction performance on jump directions, it is necessary to exploit the optimal class balancing method among all available class balancing techniques. Third, deep learning algorithms recently attract much attention in stock price prediction, and should be worthy of discussing in further study.

\section*{Acknowledgements}
This work was supported by the Natural Science Foundation of Higher Education Institutions of Jiangsu Province (17KJB120004), the Natural Science Foundation of China (71471081) and the Humanities and Social Science Fund of the Ministry of Education (17YJA790101).
\section*{References}

\biboptions{authoryear}
\bibliography{mybib}

\setcounter{equation}{0}
\renewcommand{\theequation}{A.\arabic{equation}}
\section*{Appendix A: Computation of liquidity measures}
The return $r$ and cumulative return $R$ are two types of returns: the former is the return within the interval, and the latter is the intraday cumulative return to the end of the interval. Let $p_{t,0}$, $p_{t,n}$ be the open price and close price on day $t$. In the $i$th interval of day $t$, let $p_{t, i}$ be the last record of the striking price. Then, $r_{t,i}$ and $R_{t,i}$ are defined as:
\begin{equation}
r_{t,i} = ln(p_{t,i})-ln(p_{t,i-1}),
\end{equation}

\begin{equation}
R_{t,i} = ln(p_{t,i})-ln(p_{t-1, n}).
\end{equation}

The number of trades $k$, trading volume $v$, trading size $s$, order imbalance $oi$ and depth imbalance $di$ are all volume-related liquidity measures used to evaluate trading quantity. Let $N_{t,i}$ denote the number of records. The number of trades $k_{t,i}$ and the trading volumes $v_{t,i}$ are simply the sum of trades and volumes over all the records in the specific interval:
\begin{equation}
k_{t,i} = \sum_{j=1}^{N_{t,i}}k_j,
\end{equation}
\begin{equation}
v_{t,i} = \sum_{j=1}^{N_{t,i}}v_j.
\end{equation}

The trading size $s_{t,i}$ is the average volume per trade
\begin{equation}
s_{t,i} = v_{t,i}/k_{t,i},
\end{equation}
where $v_j$ and $k_j$ be the trading volume and the number of trades of the $j$th record.

The order imbalance $oi$ computes the excess volume of buyer-initiated or seller-initiated orders. Let $I\{sign=1\}$ and $I\{sign= -1\}$ denote buyer-initiated and seller-initiated trading, respectively. The order imbalance $oi_{t,i}$ is defined as:
\begin{equation}
oi_{t,i} = \frac{2}{v_{t,i}}\left(\sum_{j=1}^{N_{t,i}}v_jI\{sign=1\}-\sum_{j=1}^{N_{t,i}}v_jI\{sign=-1\}\right).
\end{equation}

The depth imbalance $di$ represents the volume difference between the best bid and ask quotes. Let $va_j^{(1)}$ and $vb_j^{(1)}$ be the best waiting bid and ask volume of the $j$th record in the $i$th interval of day $t$, and let $pa_j^{(1)}$ and $pb_j^{(1)}$ be the best ask and bid prices of the same record. $DI_{t,i}$ is calculated by
\begin{equation}
di_{t,i} = 2\left(\sum_{j=1}^{N_{t,i}}va_j^{(1)}-\sum_{j=1}^{N_{t,i}}vb_j^{(1)}\right)\left(\sum_{j=1}^{N_{t,i}}va_j^{(1)}+\sum_{j=1}^{N_{t,i}}vb_j^{(1)}\right)^{-1},
\end{equation}

The quoted spread $qs$ and effective spread $es$ evaluate the trading cost, which are two types of volume-weighted spreads on the best ask and bid quotes:
\begin{equation}
qs_{t,i} = \frac{1}{v_{t,i}}\sum_{j=1}^{N_{t,i}}qs_jv_j,
\end{equation}
\begin{equation}
es_{t,i} = \frac{1}{v_{t,i}}\sum_{j=1}^{N_{t,i}}es_jv_j,
\end{equation}
where
\begin{equation}
qs_j = 2\left(pa_j^{(1)}-pb_j^{(1)}\right)\left(pa_j^{(1)}+pb_j^{(1)}\right)^{-1}
\end{equation}
and
\begin{equation}
es_j = \left|4p_j-2\left(pa_j^{(1)}+pb_j^{(1)}\right)\right|\left(pa_j^{(1)}+pb_j^{(1)}\right)^{-1}
\end{equation}
are the quoted spread and effective spread on the $j$th record.

The realized volatility $rv_{t,i}$, defined as the sum of squared returns within the interval, is an accurate estimate of the volatility:
\begin{equation}
rv_{t,i} = \sum_{j=1}^{N_{t,i}}rr^2_j,
\end{equation}
where $rr_j = ln\left(pa_j^{(1)}+pb_j^{(1)}\right)-ln\left(pa_{j-1}^{(1)}+pb_{j-1}^{(1)}\right)$
is the mid-price return of the best ask and bid quotes on the $j$th record.

\setcounter{equation}{0}
\renewcommand{\theequation}{B.\arabic{equation}}
\section*{Appendix B: Computation of technical indicators}
Let $O_k, H_k, L_k, C_k, V_k$ be the open price, highest price, lowest price, close price and trading volume of interval $k$.

Two types of rate of change in terms of price and volume, are useful to identify the trend of price change:
\begin{equation}
PROC(q)_k = \frac{C_k-C_{k-q}}{C_{k-q}},
\end{equation}
\begin{equation}
VROC(q)_k = \frac{V_k-V_{k-q}}{V_{k-q}}.
\end{equation}

The moving average $MA$ and exponential moving average $EMA$ are two types of indicators to determine potential changes in price series. $MA(q)_k$ and $EMA(q)_k$ of prices over $q$ lagged periods are defined by
\begin{equation}
MA(q)_k = \frac{\sum_{i=1}^{q}C_{k-i+1}}{q},
\end{equation}
\begin{equation}
EMA(q)_k = \frac{2}{q+1}C_k+\frac{q-1}{q+1}EMA(q)_{k-1}.
\end{equation}

With respect to the two types of moving averages, two types of biases ($BIAS(q)_k$, $EBIAS(q)_k$) and two types of oscillators ($OSCP(q)_k$, $EOSCP(q)_k)$) are defined to evaluate the deviation of the current market conditions to historical levels:
\begin{equation}
BIAS(q)_k = \frac{C_k-MA(q)_k}{MA(q)_k},
\end{equation}
\begin{equation}
EBIAS(q)_k = \frac{C_k-EMA(q)_k}{EMA(q)_k}.
\end{equation}
\begin{equation}
OSCP(q)_k = \frac{MA(q)_k - MA(q)_{k-1}}{MA(q)_{k-1}},
\end{equation}
\begin{equation}
EOSCP(q)_k = \frac{EMA(q)_k - EMA(q)_{k-1}}{EMA(q)_{k-1}}.
\end{equation}

The stochastic indicators, such as, fast percentK $fK$, fast percentD $fD$ and slow percentD $sD$, measure where the current price is in relation to the recent trading range. These indicators are good at identifying oversold or overbought market situations. Define $HH(q)_k = \max\{H_k, H_{k-1}, \cdots, H_{k-q+1}\}$ and $LL(q)_k =\min\{L_k, L_{k-1},$ $\cdots, L_{k-n+1}\}$. This information is used by the computation of stochastic \%K $fK(q)_k$:
\begin{equation}
fK(q)_k = \frac{C_k-LL(q)_k}{HH(q)_k-LL(q)_k},
\end{equation}
Fast stochastic \%D $fD(q)_k$ and slow stochastic \%D $sD(q)_k$ are then computed by:
\begin{equation}
fD(q)_k = \frac{fK(q)_{k-2}+fK(q)_{t-1}+fK(q)_k}{3},
\end{equation}
\begin{equation}
sD(q)_k = \frac{fD(q)_{k-2}+fD(q)_{t-1}+fD(q)_k}{3}.
\end{equation}

Accumulation/distribution oscillator $ADO_k$, true range $TR_k$ and commodity channel index use various price information to measure the volatility of prices:
\begin{equation}
ADO_k = \frac{(H_k-O_k)-(C_k-L_k)}{2\times(H_k-L_k)},
\end{equation}
\begin{equation}
TR_k = \max\{H_k-L_k, L_k-C_{k-1}, H_k-C_{k-1}\},
\end{equation}
\begin{equation}
CCI(q)_k = \frac{M_k - SM(q)_k}{0.015G(q)_k},
\end{equation}
where
$M_k = (H_k+L_k+C_k)/3$, $SM(q)_k = (\sum_{i=1}^{q} M_{k-i+1})/q$ and $G(q)_k = (\sum_{i=1}^{q}|M_{k-i+1}-SM(q)_k|)/q$.

Some indicators, such as price and volume trend $PVT$, on-balance volume $OBV$, negative volume index $NVI$ and positive volume index $PVI$, combine information about both the price and volume series. $PVT$ confirms the strength of price trends:
\begin{equation}
PVT_k = \frac{C_k-C_{k-1}}{C_{k-1}}\times V_k,
\end{equation}
$OBV$ analyses how the volume flow is influenced by price:
\begin{equation}
OBV_k = OBV_{k-1}+\left\{\begin{array}{rll}
V_k,       &{if\ C_k > C_{k-1}},\\
0,    	&{if\ C_k=C_{k-1}},\\
-V_k,     &{else}.\\
\end{array} \right.
\end{equation}
$NVI$ and $PVI$ reveal how the price is influence by the volume:
\begin{equation}
NVI_k =NVI_{k-1}\times \left\{\begin{array}{rll}
1 ,       &{if\ V_k \geq V_{k-1}},\\
\frac{C_k}{C_{k-1}},    	&{else}.\\
\end{array} \right.
\end{equation}
\begin{equation}
PVI_k = PVI_{k-1} \times \left\{\begin{array}{rll}
1,       &{if\ V_k \leq V_{k-1}},\\
\frac{C_k}{C_{k-1}},    	&{else}.\\
\end{array} \right.
\end{equation}

\end{document}